\documentclass[aps,prl,twocolumn,showpacs,superscriptaddress,floatfix,notitlepage]{revtex4-1}

\usepackage{float}
\usepackage{tabularx}
\usepackage{amsmath}
\usepackage{adjustbox}
\usepackage{booktabs}
\usepackage[x11names]{xcolor}
\usepackage{fancyhdr, amssymb, cancel, amsmath, graphicx}
\usepackage{isomath}
\usepackage[colorlinks=true,hyperindex=true, linktocpage=true]{hyperref}
\usepackage[capitalize]{cleveref}

\usepackage{subcaption}
\usepackage{ragged2e}
\DeclareCaptionJustification{justified}{\justifying}
\begin{document}

\title{Viscometry of electron fluids from symmetry}

\author{Caleb Q. Cook}
\email{calebqcook@gmail.com}
\affiliation{Department of Physics, Stanford University, Stanford CA 94305, USA} 

\author{Andrew Lucas} 
\email{andrew.j.lucas@colorado.edu}
\affiliation{Department of Physics, University of Colorado, Boulder CO 80309, USA}
\affiliation{Center for Theory of Quantum Matter, University of Colorado, Boulder CO 80309, USA}

\begin{abstract}
When electrons flow as a viscous fluid in anisotropic metals, the reduced symmetry can lead to exotic viscosity tensors with many additional, non-standard components. We present a viscometry technique that can in principle measure the multiple dissipative viscosities allowed in isotropic and anisotropic fluids alike. By applying representation theory to exploit the intrinsic symmetry of the fluid, our viscometry is also exceptionally robust to both boundary complications and ballistic effects. We present the technique via the illustrative example of dihedral symmetry, relevant in this context as the point symmetry of 2D crystals. Finally, we propose a present-day realizable experiment for detecting, in a metal, a novel hydrodynamic phenomenon: the presence of rotational dissipation in an otherwise-isotropic fluid. 
\end{abstract}

\maketitle

\textit{Introduction---}Hydrodynamics models the transport of conserved quantities, such as charge or energy, over large length- and time-scales.  In ultra-pure low-temperature metals, electronic momentum can also be approximately conserved, if the collisions that conserve momentum are much faster than those that relax it (e.g. off impurities or via umklapp) \cite{lucasreview17}. In these viscous electron fluids, hydrodynamic effects can give rise to exotic transport phenomena, such as decreasing resistance with increasing temperature (Gurzhi effect) \cite{gurzhi} and superballistic constriction flow \cite{levitov1607}.  

Theorized for many decades, electron hydrodynamics has in recent years garnered compelling experimental evidence \cite{molenkamp,crossno,bandurin,kumar2017superballistic,bandurin18,bakarov,imaging1,imaging2,imaging3}. The earliest discoveries of electron hydrodynamics took place in GaAs \cite{molenkamp}, monolayer graphene \cite{crossno}, and bilayer graphene \cite{bandurin}.  At low (but non-zero) charge density, these are all isotropic Fermi liquids well-described by Galilean-invariant, textbook hydrodynamics \cite{landau}. For the electron fluid in graphene, the shear viscosity -- the sole dominant viscosity in this isotropic Fermi liquid -- has been both calculated \cite{polini1506,NarozhnyVisc} and indirectly measured in experiment \cite{bandurin,kumar2017superballistic,imaging2}.

Metals are generically anisotropic, however, as the presence of a crystalline lattice explicitly breaks rotational symmetry. Indeed, experiments and \textit{ab initio} calculations have recently suggested hydrodynamics might apply in less symmetric metals, e.g. $\mathrm{WP}_2$ \cite{gooth2018thermal}, $\mathrm{PtSn}_4$ \cite{fu2018thermoelectric},
$\mathrm{MoP}$ \cite{mop2019}, $\mathrm{WTe}_2$ \cite{vool2020imaging}. In such cases, anisotropy leads to a number of novel phenomena \cite{varnavides2020electron}, including rotational viscosity \cite{cook2019} and intrinsic Hall viscosity \cite{Toshio_2020}. Such viscosities are inaccessible to current experiments, however, as existing methods (non-local resistances \cite{polini,levitovhydro}, constriction conductances \cite{levitov1607}, AC phenomena \cite{CorbinoVisc}, current imaging \cite{imaging1,imaging2,imaging3}, channel flows \cite{link}, and heat transport \cite{vignale,gooth2018thermal,heatTransport1,heatTransport2}) \textit{(i)} are not robust to boundary and ballistic effects, and \textit{(ii)} cannot distinguish all the symmetry-allowed viscosities that will generically appear. 

Here, we present a multi-terminal device, robust to both boundary complications and ballistic effects, that can measure the multiple  dissipative viscosity components allowed in isotropic \emph{and} anisotropic fluids, all on a single sample. Our viscometry relies on the representation theory of point groups, from which we devise boundary conditions that isolate viscosities via symmetry-constrained heating. Our technique is also uniquely capable detecting a ``smoking gun" signal of a novel hydrodynamic phenomenon: the isolated emergence of rotational viscosity $\eta_\circ$ in an ``otherwise isotropic" fluid \cite{cook2019}.

Strikingly, rotational viscosity $\eta_\circ$ gives viscous dissipation \emph{even under rigid rotations of a fluid}, which is forbidden by angular momentum conservation in isotropic fluids, but generically allowed in anisotropic fluids. For hexagonal fluids in particular, $\eta_\circ$ emerges in a novel and isolated way \cite{cook2019}, alongside only the standard, isotropic shear and bulk viscosities. Hexagonal electron fluids therefore provide a highly novel setting for finding $\eta_\circ$, with possible candidate materials including $\mathrm{PdCoO}_2$ \cite{mackenzie}, $\mathrm{Na}{\mathrm{Sn}}_{2}{\mathrm{As}}_{2}$ \cite{wang2020goniopolar}, and $\mathrm{ABA}$-trilayer graphene \cite{Zibrov_2018}. Finally, we argue that our viscometry proposed here is in fact the \emph{only} feasible way of discovering $\eta_\circ$ in an electron fluid.

In what follows, we describe our viscometry via the illustrative example of 2D fluids of dihedral point symmetry. However, our approach extends naturally to fluids of higher dimension and/or differing point symmetry.

\textit{Dihedral hydrodynamics---}The dihedral group $\mathsf{D}_{2M}$ is the $2M$-element group of symmetries of the regular $M$-gon. As an abstract group, $\mathsf{D}_{2M}$ is generated by its elements $\rho$, a ($2\pi/M$)-rotation about the $M$-gon center, and $r$, a reflection through a fixed axis containing the $M$-gon center, with $\rho r\rho = r$. We also take $\mathsf{D}_\infty=\mathsf{O}(2)$ to be the group of symmetries of the circle, which includes rotations of arbitrary angle. By the crystallographic restriction theorem \cite{ashcroft}, the paradigmatic 2D electron fluids are those of $M\in\{2,3,4,6\}$ dihedral point symmetry.

In Newtonian fluids (appropriate for the linear response regime \cite{lucasreview17}), viscous stresses $\tau_{ij}=-\eta_{ijkl}\partial_k v_l$ arise linearly in response to velocity gradients $\partial_k v_l$, with proportionality given by the viscosity tensor $\eta_{ijkl}$. In the Supplemental Material (SM), we show that any $\mathsf{D}_{2M}$-invariant viscosity tensor must take the form
\begin{widetext}
\begin{equation}
\label{eq:viscTensor}
\eta_{ijkl}=\begin{cases}
\eta(\sigma_{ij}^{x}\sigma_{kl}^{x}+\sigma_{ij}^{z}\sigma_{jk}^{z})+\zeta(\delta_{ij}\delta_{kl}), & M=\infty\\
\eta(\sigma_{ij}^{x}\sigma_{kl}^{x}+\sigma_{ij}^{z}\sigma_{jk}^{z})+\zeta(\delta_{ij}\delta_{kl})+\eta_{\circ}(\epsilon_{ij}\epsilon_{kl}), & M\in\{3\}\cup[5,\infty)\\
\eta_{\times}(\sigma_{ij}^{x}\sigma_{kl}^{x})+\eta_{+}(\sigma_{ij}^{z}\sigma_{jk}^{z})+\zeta(\delta_{ij}\delta_{kl})+\eta_{\circ}(\epsilon_{ij}\epsilon_{kl}), & M=4
\end{cases}
\end{equation}
\end{widetext}
where $\epsilon$ is the Levi-Civita symbol and $\sigma^a$ are Pauli matrices. We have excluded in \cref{eq:viscTensor} only the $M=2$ viscosity tensor; in such $\mathsf{D}_4$ fluids, one has eight allowed viscosities, not all of which are isolated by our viscometry due to the exceptionally-low symmetry of $\mathsf{D}_4$. We therefore relegate discussion of this singular case to SM.

\begin{figure}[t!]
\centering
    \includegraphics[width=0.185\textwidth]{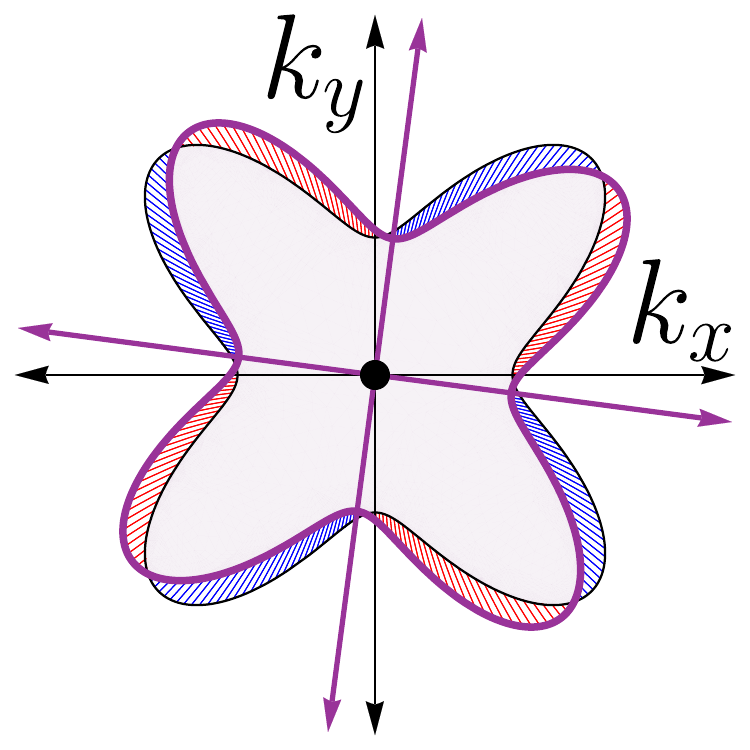}
    \caption{
   Illustration of the origin of rotational viscosity in electron fluids. When an anisotropic Fermi surface (black) is rotated (dark purple), quasiparticle excitations (red/blue) are generated. In the hydrodynamic limit, such rigid rotations are opposed by a dissipative rotational viscosity $\eta_\circ$ \cite{cook2019}. Note that this Fermi surface has $\mathsf{D}_8$ symmetry.}
    \label{fig:Micro}
\end{figure}

We emphasize that the presence of rotational viscosity $\eta_\circ$ in \cref{eq:viscTensor} does not rely on electrons or dihedral symmetry: it is universal to anisotropic fluids. The lack of rotational symmetry allows the stress tensor to have a non-vanishing antisymmetric component $\epsilon_{ij}\tau_{ij}\neq0$, which in the hydrodynamics must couple to the strain tensor component $\epsilon_{ij}\partial_i v_j=\nabla\times\mathbf{v}$ of the same symmetry (i.e. the vorticity); this generic coupling is $\eta_\circ$. \cref{fig:Micro} illustrates the microscopic origin of $\eta_\circ$ in anisotropic electron fluids.

The remaining viscosities appearing in \cref{eq:viscTensor} can be understood as follows: bulk viscosity $\zeta$ \cite{bulkFN} couples the trace of the stress tensor to the fluid expansion $\nabla\cdot\mathbf{v}$, plus viscosity $\eta_+$ couples the stress $(\tau_{xx}-\tau_{yy})$ along the axes of the crystal to the strain $(\partial_x v_x - \partial_y v_y)$, and cross viscosity $\eta_\times$ couples stress and strain at 45$^\circ$ to the crystal axes. Equating plus and cross viscosities $\eta_+,\eta_\times\to\eta$ in the $\mathsf{D}_8$ tensor ($M=4$) gives the $\mathsf{D}_{12}$ tensor ($M=6$), and further taking $\eta_\circ\to0$ in the $\mathsf{D}_{12}$ tensor gives the isotropic tensor ($M=\infty$). We therefore discuss dihedral viscosities without further loss of generality by henceforth assuming the $\mathsf{D}_8$ case.

We now turn to the linearized (i.e. assuming Stokes flow \cite{lucasreview17,landau}) hydrodynamics. For $\mathsf{D}_8$ fluids, the hydrodynamic equations are the following pair of approximate conservation laws:
\begin{subequations}
\begin{align}
    \partial_{t}\rho	&=-\partial_{i}\left(\rho_{0}v_{i}-D\partial_{i}\rho\right),
    \label{eq:introNScon}
    \\
    \rho_{0}\partial_{t}v_{i}	&=-c^2\partial_{i}\rho-\rho_0\Gamma v_i+\eta_{jikl}\partial_{j}\partial_{k}v_{l},
    \label{eq:introNSmom}
\end{align}
\label{eq:introNS}
\end{subequations}
where $\rho$ ($\rho_0$) is the (equilibrium) fluid density, $c$ the electronic speed of sound, and $\Gamma$ is the rate of momentum-relaxing collisions. \cref{eq:introNScon} describes the local conservation of density $\rho$, with an associated conserved current $J_i=\rho_0 v_i - D\partial_i \rho$. The current $J_i$ has a convective contribution from the fluid momentum $\rho_0 v_i$ and a diffusive contribution $-D\partial_i\rho$, with $D$ the incoherent diffusion constant \cite{cook2019,hartnoll1}. \cref{eq:introNSmom} describes the approximate conservation of fluid momentum $\rho_0 v_i$ in the presence of viscous $-\partial_j\tau_{ji}$ and ohmic $-\rho_0\Gamma v_i$ forces.   

One may in principle append to \cref{eq:introNS} a third conservation law for energy. At $\rho_0\neq 0$, this complication does not qualitatively modify the dynamics of homogeneous electron fluids \cite{lucasreview17}. At $\rho_0=0$ (e.g. the Dirac fluid of charge-neutral graphene), the energy density $\epsilon$ couples to velocity $v_i$ in an analogous way to charge density $\rho$ in \cref{eq:introNS}. Due to this analogy we focus on the $\rho_0\neq0$ case, but our results are generalizable to Dirac fluids.

We now restrict to static flows $\partial_t=0$, so that the left-hand-side of \cref{eq:introNS} vanishes. We can then automatically satisfy the resulting divergence-free condition on $J_i$ \labelcref{eq:introNScon} by writing the current in terms of a stream function: $J_{i}\equiv\rho_{0}\epsilon_{ij}\partial_{j}\psi\implies v_{i}=(D/\rho_{0})\partial_{i}\rho+\epsilon_{ij}\partial_{j}\psi.$ Using this stream function $\psi$, we eliminate density $\rho$ from the (static) momentum equation \labelcref{eq:introNSmom} and, neglecting terms of order $\eta D\partial^{2}\psi\sim\left(\ell_{\text{ee}}\partial\right)^{2}$, we find that the stream function satisfies the generalized biharmonic equation
\begin{equation}
    \overline{\nabla}^{4}\psi
    =
    \left(\frac{w}{\lambda}\right)^2\overline{\nabla}^{2}\psi
    +
    \delta\left[\left(\partial_{\overline{x}}^{2}-\partial_{\overline{y}}^{2}\right)^{2}-\left(2\partial_{\overline{x}}\partial_{\overline{y}}\right)^{2}\right]\psi,
    \label{eq:bi2}
\end{equation}
where we have introduced the parameters
\begin{equation}
    \lambda	=\sqrt{\frac{2\eta_{\circ}+\eta_{+}+\eta_{\times}}{2\rho_{0}\Gamma}},
    \;
    \delta	=\frac{\eta_{+}-\eta_{\times}}{2\eta_{\circ}+\eta_{+}+\eta_{\times}},
    \label{eq:params}
\end{equation}
and non-dimensionalized all lengths $( \overline{x},\overline{y})\equiv(x,y)/w$, $\overline{\nabla}\equiv\langle\partial_{\overline{x}},\partial_{\overline{y}}\rangle$, using an assumed measurement lengthscale $w$ (which will later characterize the size of our viscometer). Using an assumed solution $\psi$ of the generalized biharmonic \labelcref{eq:bi2}, we solve for $\partial_i \rho$ in \cref{eq:introNSmom}, which tells us that (away from $\rho_0=0$) the current $J_i\approx \rho_0 v_i$ is approximately coherent at this order \cite{incFN}. Substituting this result into the stream function relation, we find that the fluid is approximately incompressible: $v_i \approx \epsilon_{ij}\partial_j \psi$.

The parameter $\lambda$ \labelcref{eq:params} is known as the \emph{Gurzhi length} and characterizes the length-scale past which momentum-relaxing effects begin to dominate viscous effects \cite{lucasreview17}. The dimensionless parameter $\delta$ \labelcref{eq:params} characterizes the degree of square anisotropy in the fluid and must lie in the interval $\delta\in[-1,1]$. The transformation $\delta\to-\delta$ corresponds to a rotation of the crystal coordinates by $45^\circ$, and $\delta=0$ implies $\eta_+=\eta_\times$ (no square anisotropy in the fluid).

\textit{Dihedral viscometry---}Our dihedral viscometer is a square $(x,y) \in\left[-w/2,w/2\right]^{2}$, with current $J_i\approx \rho_0 v_i$ boundary conditions consisting of $8$ contacts, each of width $a$, on its perimeter. Contacts are placed in pairs symmetrically about the midpoint of each edge, separated from each other by a tunable spacing $d$. A total current $I_0$ is either injected or drained at each contact, with the configuration of the viscometer determined by these choices. For concreteness, we take box function contacts \cite{boxFN}, and no-slip $v_i=0$ at the boundary away from contacts, in all numerical calculations (though our main results are unaffected by such details). 

Our viscometry functions by exploiting the spatial symmetry of the dissipation generated in the fluid. The viscous dissipation is best understood via the irreducible symmetries of the $\mathsf{D}_8$-invariant viscosity tensor, which we now outline; see SM for details.

Informally, a \emph{group representation} \cite{tung} allows a group to act on a vector space, by assigning group elements to matrices in a way that is consistent with the underlying group multiplication. For finite groups and complex vector spaces, any such representation can be decomposed into a sum of elementary, ``building-block" representations, known as \emph{irreducible representations} (irreps). The dihedral group $\mathsf{D}_{8}$ has five irreps: four $1$-dimensional representations $U_{0, 2}^{\pm}$ (the superscript denotes reflection parity, $U_{k}^{\pm}\left(r\right)=\pm1$, and the subscript denotes rotation parity, $U_{k}^{\pm}\left(\rho\right)=\mathrm{i}^k$) and one $2$-dimensional vector representation $R_1$ \cite{cook2019, tung}. These irreps label the five irreducible ways a mathematical object can self-consistently transform under reflection and 4-fold rotation. The irreps of $\mathsf{D}_8$ and their realizations as current boundary conditions on a square are summarized in \cref{table:irreps}.

\begin{table}[t!]
\includegraphics[width=.475\textwidth]{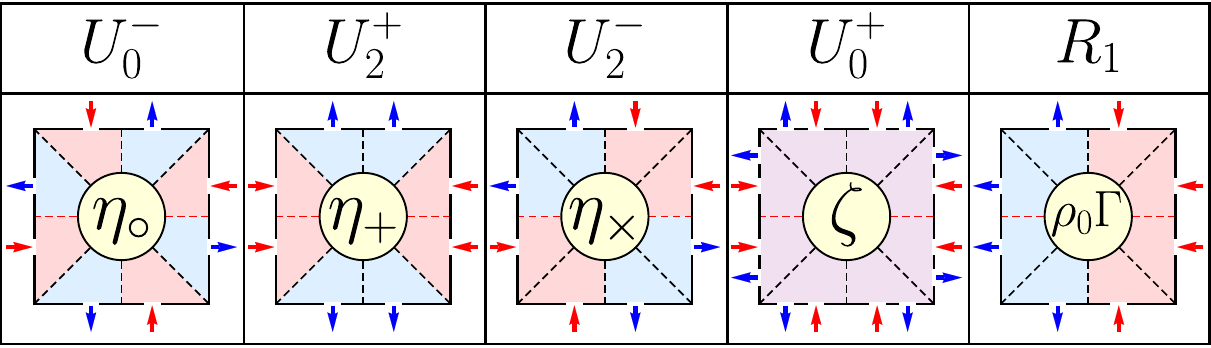}
\caption[]{
\textit{First row}: The five irreducible representations of $\mathsf{D}_8$. \textit{Second row}: Current boundary conditions (blue/red arrows) of matching $\mathsf{D}_8$-symmetry, indicated by colored wedges. Symmetry restricts heat \labelcref{eq:velocityHeat} at the square center to \emph{only} a single dissipative coefficient (yellow disk). Note that the representation $U_0^+$ requires more than 8 contacts in order to satisfy charge conservation.
}
\label{table:irreps}
\end{table}

Particularly relevant for viscometry is the $4$-dimensional vector space $\mathcal{T}_2$ of rank-2 tensors, as the velocity strain tensor is an element of this space: $\partial_i v_j\in \mathcal{T}_2$.  The viscosity tensor $\eta_{ij,kl}\equiv \eta_{ijkl}$ then acts linearly on $\mathcal{T}_2$ as a $4\times 4$ matrix by index contraction.  Since the viscosity tensor is $\mathsf{D}_8$-invariant, Schur's lemma \cite{tung} implies that $\eta_{ij,kl}$ must act proportionally to the identity on each $\mathsf{D}_8$-invariant subspace of $\mathcal{T}_2$. We illustrate this result by expressing the heat that is generated through viscous dissipation, 
$W_{\text{visc}}=(\partial_{i}v_{j})\eta_{ij,kl}(\partial_{k}v_{l})$, as
\begin{equation}
\begin{aligned}
    W_{\text{visc}}  
    =\; 
    &\eta_{\circ}\left(\epsilon_{ij}\partial_{i}v_{j}\right)^{2}
    +
    \eta_{+}(\sigma_{ij}^{z}\partial_{i}v_{j})^{2}\\
    & +\eta_{\times}(\sigma_{ij}^{x}\partial_{i}v_{j})^{2}+\zeta(\delta_{ij}\partial_{i}v_{j})^{2},
\label{eq:velocityHeat}
\end{aligned}
\end{equation}
where each term in \cref{eq:velocityHeat} represents a projection of $\partial_i v_j$ into a given 1-dimensional $\mathsf{D}_8$-invariant subspace of $\mathcal{T}_2$, corresponding to a $1$-dimensional irrep of $\mathsf{D}_8$.

Note that the total \cite{totalFN} heat $W=W_\text{visc}+W_\text{ohm}$ generated by the fluid flow also contains an ohmic contribution $W_\text{ohm}=\rho_0\Gamma v_i^2$. Even though $\rho_0\Gamma$ is not a component of the viscosity tensor, the fluid velocity $v_i$ nevertheless transforms according to the remaining vector irrep $R_1$, conveniently completing our correspondence between $\mathsf{D}_8$ irreps and dissipative coefficients in \cref{table:irreps}.

Importantly, both the center of the square \emph{and} its boundary are mapped to themselves under any $\mathsf{D}_8$ symmetry transformation.  Thus the center strain tensor $\left.\left(\partial_{i}v_{j}\right)\right|_{\mathbf{r}=\mathbf{0}}$ and center velocity $v_i(\mathbf{0})$ must have the same $\mathsf{D}_8$ symmetry as the square boundary. This implies that we can selectively isolate at the square center each of the $5$ terms in the heat decomposition $W=W_\text{visc}+W_\text{ohm}$ by choosing boundary conditions corresponding to each of the $5$ irreps of $\mathsf{D}_8$. 

\begin{figure}[t!]
\centering
\includegraphics[width=.5\textwidth]{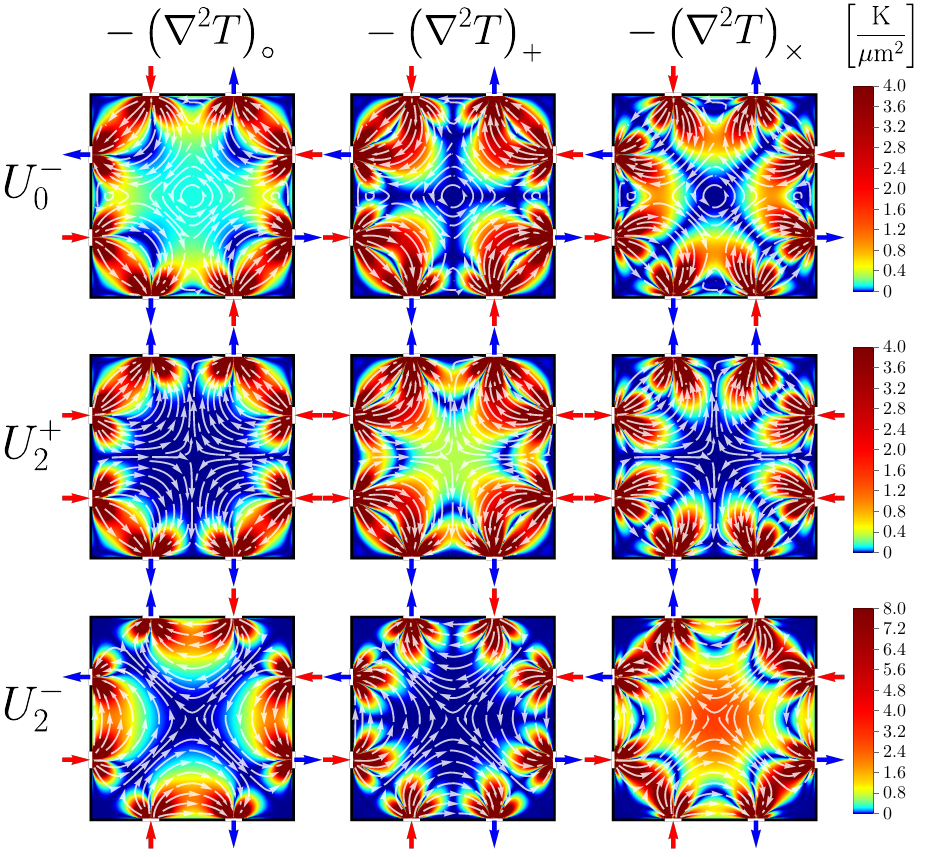}
\caption{
Flows numerically solving \cref{eq:bi2} in our viscometer with $w=1\text{ }\mu\text{m}$, $I_0 = 100\text{ }\mu\text{A}$, $d/w=0.41$, $a/w=0.05$, $\delta=0$, and $\lambda/w=\infty$. Rows specify $\mathsf{D}_8$-irreducible boundary conditions, and columns the temperature variation $-(\nabla^2 T)_\alpha$ sourced solely by $\eta_\alpha$-dissipation. Symmetry restricts center heating to only the diagonal plots. In giving an order-of-magnitude estimate for the scale of heating, we have taken relevant physical parameters from hydrodynamic electrons in monolayer graphene \cite{bandurin,kumar2017superballistic}; see SM. Temperature variations of this magnitude are detectable with existing local thermometers \cite{zhang2017anomalous,diamondTempSensing}.
}
\label{fig:heatTable}
\end{figure}

The above considerations are summarized in \cref{table:irreps}. A numerical demonstration of isolated $\eta_\circ$, $\eta_+$, and $\eta_\times$ heating is given in \cref{fig:heatTable} (see SM for additional flow plots). In SM, we further show that our result does not fundamentally rely on hydrodynamics; across the \emph{entire} ballistic-to-hydrodynamic crossover, our symmetry-based ``viscometer" continues to isolate dissipation channels according to their symmetry.

The isolated center heat $W_\mathbf{0}=\eta_\alpha (\partial v_\alpha)^2_\mathbf{0}$ generated solely by the viscosity $\eta_\alpha$ sources a Poisson equation \cite{crossno}
\begin{equation}
    W=-\kappa \nabla^2 T
\label{eq:poissonW}   
\end{equation}
for temperature $T$, with $\kappa$ the electronic thermal conductivity. If one is able to measure both the center temperature variation $(\nabla^2 T)_\mathbf{0}$ (e.g. by local thermometry \cite{zhang2017anomalous,diamondTempSensing})  and center strain component $(\partial v_\alpha)_{\mathbf{0}}$ (e.g. by flow imaging \cite{imaging1,imaging2,imaging3}), then $\eta_\alpha = -\kappa(\nabla^2 T)_\mathbf{0}/(\partial v_\alpha)^2_\mathbf{0}$ can be determined. Alternatively, if one uses \emph{only} local thermometry, one may still estimate $(\partial v_\alpha)_{\mathbf{0}}$ -- and hence $\eta_\alpha$ -- by mapping out heating patterns $W(x,y)$ via \cref{eq:poissonW} and comparing against numerical simulations.

\begin{figure}[t!]
    \centering
    \includegraphics[width=0.5\textwidth]{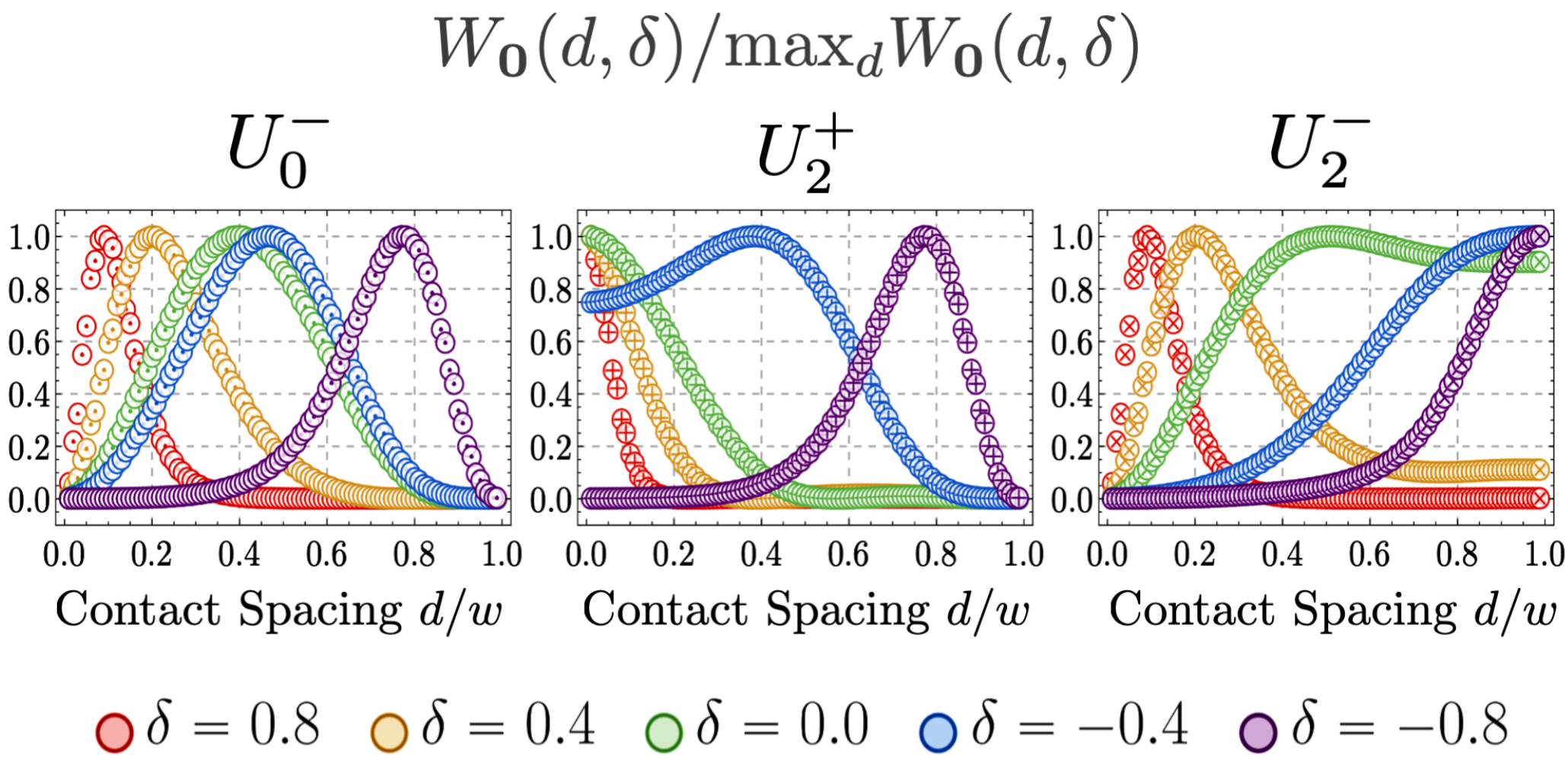}
    \caption{
    Viscometer center heat $W_\mathbf{0}$, numerically determined from \cref{eq:bi2}, as a function of boundary condition irrep., contact spacing $d$, and anisotropy $\delta$, for $a/w=0.01$ and $\lambda/w=\infty$. Each curve is normalized by its max value. The uniqueness of these curves should allow for experimental determination of $\delta$. Although momentum-relaxation is neglected in these $\lambda/w=\infty$ plots, we find that the shape of these curves, and hence their utility in determining $\delta$, is extremely insensitive to decreasing $\lambda$ (increasing $\Gamma$); see SM.
    }
    \label{fig:invertHeats}
\end{figure}

Another consistency check arises by varying the viscometer geometry. Numerically solving \cref{eq:bi2} for varying contact spacing $d$, we show in \cref{fig:invertHeats} how the anisotropy $\delta$ can be determined experimentally. The center heat $W_\mathbf{0}(d)$ (as a function of contact spacing $d$) varies uniquely with anisotropy $\delta$, allowing for computation of the latter. In fact, we show in SM how $\delta$ may be determined from as few as $2$ contact spacings and $2$ boundary configurations, for $4$ total center heat measurements.

Finally, in SM we discuss how our viscometry compares against more conventional Poiseuille, channel flow methods, particularly in the $\mathsf{D}_4$ case \cite{link} where there is insufficient symmetry to isolate all viscosities via boundary conditions, as above.


\textit{Conclusions---}Even if the above procedure cannot be carried out in full, one may nevertheless \emph{detect} rotational viscosity $\eta_\circ$ by simply observing center heat in the $U_0^-$ configuration. $U_0^-$-symmetry precludes any center heat that might arise from another viscosity component, ohmic effects, incoherent currents, or even ballistic scattering (in addition to being highly suppressed in the viscous limit, ballistic center heat also has easily distinguishable scaling with viscometer size $w$; see SM). We therefore anticipate that our viscometry can enable the discovery of $\eta_\circ$ in the near future. 

We further claim that (in contrast to other dihedral viscosities) there is no feasible way to detect $\eta_\circ$ beyond the symmetry-based technique proposed here. Expanding the hexagonal viscosity tensor \labelcref{eq:viscTensor} in \cref{eq:introNSmom}, one in fact obtains the \emph{isotropic} momentum equation, but with replacements $\left\{ \eta,\zeta\right\} \to\left\{ \eta+\eta_{\circ},\zeta-\eta_{\circ}\right\}$ \cite{vectorCalcFN}. This implies that \emph{rotational viscosity does not modify bulk flow patterns}. Although exotic no-stress boundary conditions can in principle generate weakly $\eta_\circ$-dependent flows, the incomplete understanding of viscous electron boundary conditions makes it is unclear how such an experiment could be robustly carried out.

Indeed, there has been much discussion concerning the proper boundary conditions (e.g. no-slip, no-stress, generalized Robin) for viscous electron flow \cite{wagner2015boundary, kiselev2019boundary, moessner2019boundary}. Because our viscometer relies on symmetry, it conveniently side-steps any such boundary complication, so long as the boundaries are symmetrically complicated.  For example, although we assumed no-slip $v_i=0$ boundary conditions in the preceeding numerics, if no-stress or generalized Robin boundary conditions are instead required, the numerical values in \cref{fig:heatTable,fig:invertHeats} will change but the irrep decomposition of the rank-2 tensor space $\mathcal{T}_2$ will continue to guarantee isolated center heating.  

We emphasize that our viscometry extends to more general fluids. For fluids of point group symmetry $G$, one constructs a device with $G$-irreducible boundary conditions. Then the viscous heat generated at a $G$-invariant point (i.e. mapped to itself under the action of $G$) can be selectively restricted to each irreducible comnponent of the viscosity tensor, as above. Our viscometery therefore also generalizes to higher dimensions, although measuring local heating at the center of a 3D sample may be more challenging.


Finally, for fluids with broken inversion and time-reversal symmetries, additional non-dissipative tensors \cite{hallViscPRX2020,BurmistrovHallVis2019,viscTensors2020} may appear in $\eta_{ijkl}$ \labelcref{eq:viscTensor}. We compute these lower-symmetry tensors in SM, matching those found in recent work on anisotropic Hall viscosities \cite{hallViscPRX2020}. We expect our viscometry to \emph{partially} extend to such fluids, since tailored boundary conditions will be able to similarly isolate in experiment the effects of symmetry-constrained Hall viscosisties. However, while neither Hall viscosity nor $\eta_\circ$ modify the form of the Navier-Stokes equations, the Hall viscosity is, moreover, \emph{non-dissipative}. Thus, for our viscometry to prove fully applicable to Hall viscosities, an experimental signature beyond heating must first be identified.

We thank Irving Dai and David Goldhaber-Gordon for helpful discussions. CQC was supported by NSF Grant No. DMR2000987.  AL was supported by a Research Fellowship from the Alfred P. Sloan Foundation through Grant FG-2020-13795, and by the Gordon and Betty Moore Foundation's EPiQS Initiative via Grant GBMF10279.

\bibliographystyle{unsrt}
\bibliography{aniHydro}

\begin{thebibliography}{10}
\bibitem{bulkFN}
In a viscous Fermi liquid, the bulk viscosity
  $\zeta\sim\left(T/T_\text{F}\right)^4 \mu$ is strongly suppressed at low
  temperature relative to other viscosity components $\mu$ \cite{sykes} and
  therefore often neglected. Additionally, in this work, we make approximations
  which lead to an incompressible fluid
  $v_{i}\approx\epsilon_{ij}\partial_{j}\psi$ and therefore remove $\zeta$
  entirely from the dynamics of the fluid. However, from symmetry
  considerations alone, we nevertheless propose a device (\cref{fig:sq16_bulk})
  which isolates the dissipative effects of $\zeta$ and could potentially
  thereby enable its measurement.
\bibitem{boxFN}
$I\left(s\right)=\pm I_{0}\text{rect}\left[\left(s\pm d/2\right)/a\right]$,
  where $\text{rect}(x)$ is defined to be $1$ for $|x|\leq 1/2$, and $0$
  otherwise.
\bibitem{vectorCalcFN}
This is a consequence of the vector calculus identity
  $\nabla\times\left(\nabla\times\mathbf{v}\right)=\nabla\left(\nabla\cdot\mathbf{v}\right)-\nabla^{2}\mathbf{v}$.
\end{thebibliography}
\onecolumngrid
\pagebreak

\begin{appendix}
\begin{center}
    {\Large \textbf{Supplementary material for \\``Viscometry of electron fluids from symmetry"}}
\end{center}

\section{Representation theory} 
\label{sec:gTheory}


\subsubsection{Dihedral groups}
We briefly summarize the representation theory of dihedral groups $\mathsf{D}_{2M}$ of degree $M$, as well as the continuous group $\mathsf{O}(2)\equiv \mathsf{D}_\infty$, which we will regard as an infinite generalization of a dihedral group. Further explanation of terminology and results presented here may be found in Appendix C of \cite{cook2019}. 

The orthogonal group $\mathsf{O}(2)$ is the continuous group of distance-preserving transformations
of the Euclidean plane. $\mathsf{O}(2)$ may be presented as:
\begin{equation} 
\mathsf{O}(2)=\left\langle r,\left\{ \rho_{\theta}\right\} _{\theta\in[0,2\pi]}\left|\right.r^{2}=\rho_{2\pi}=\rho_0=1,\rho_{\theta}\rho_{\phi}=\rho_{\theta+\phi},r\rho_{\theta}r=\rho_{-\theta}\right\rangle .
\end{equation}
The irreducible representations of the orthogonal group $\mathsf{O}(2)$ are
precisely two $1$-dimensional representations $\mathcal{U}_{0}^{\pm}$
and infinitely many $2$-dimensional representations $\mathcal{R}_{k}$
labeled by positive integers $k\in\mathbb{N}$. They are defined by:
\begin{subequations}
\begin{align}
    \mathcal{U}_{0}^{\pm}(\rho_{\theta})&=1,\\
\mathcal{U}_{0}^{\pm}(r)&=\pm1, \\
\mathcal{R}_{k}(\rho_{\theta})&=\left[\begin{array}{cc}
\cos\left(k\theta\right) & \sin\left(k\theta\right)\\
-\sin\left(k\theta\right) & \cos\left(k\theta\right)
\end{array}\right],\\
\mathcal{\mathcal{R}}_{k}(r)&=\left[\begin{array}{cc}
1 & 0\\
0 & -1
\end{array}\right].
\end{align}
\label{eq:O2irreps}
\end{subequations}
Tensor products of irreducible representations of $\mathrm{O}(2)$ decompose into
direct sums of said irreducible representations according to the following rules: 
\begin{subequations}
\label{eq:LRO2}
\begin{align}
    \mathcal{U}_{0}^{\eta}\otimes\mathcal{U}_{0}^{\zeta} &= \mathcal{U}_{0}^{\eta\cdot\zeta},
    \\
    \mathcal{U}_{0}^{\pm}\otimes\mathcal{R}_{k} &= \mathcal{R}_{k},
    \\
    \mathcal{R}_{k}\otimes\mathcal{R}_{l} &= \mathcal{R}_{|k-l|}\oplus\mathcal{R}_{k+l},
\end{align}
\end{subequations}
where in the last decomposition we have defined the (reducible) representation 
\begin{equation}
\mathcal{R}_{0}\equiv\mathcal{U}_{0}^{+}\oplus\mathcal{U}_{0}^{-}.
\end{equation}

The dihedral group $\mathsf{D}_{2M}$ of order $2M$ and degree $M$ is the group of planar symmetries of a regular $M$-gon. $\mathsf{D}_{2M}$ may be presented as
\begin{equation}
    \mathsf{D}_{2M}=\left\langle r,\rho\left|\right.r^{2}=\rho^{M}=1,r\rho r=\rho^{-1}\right\rangle.
\end{equation}
Note that $\mathsf{D}_{2M}$ is a subgroup of $\mathsf{O}(2)$ for all degree $M$. 

For even degree $M$, the irreducible representations of the dihedral group
$\mathsf{D}_{2M}$ are precisely $4$ one-dimensional representations $U_{0}^{\pm},U_{M/2}^{\pm}$
and $\left(M/2-1\right)$ two-dimensional representations $R_{k}$,
with $k=1,\ldots,(M/2-1)$. They are defined by:
\begin{subequations}
\begin{align}
U_k^\pm(\rho) &= (-1)^{1-\delta_{k0}}, 
\\
U_{k}^{\pm}(r) &=\pm1, 
\\
R_k(\rho) &= \left[\begin{array}{cc}
\cos\left(k\theta_{M}\right) & \sin\left(k\theta_{M}\right)\\
-\sin\left(k\theta_{M}\right) & \cos\left(k\theta_{M}\right)
\end{array}\right]
\\
R_{k}(r)&=\left[\begin{array}{cc}
1 & 0\\
0 & -1
\end{array}\right]
\end{align}
\label{eq:DihIrreps}
\end{subequations}
where $\theta_M\equiv2\pi/M$. 

For odd degree $M$, the irreducible representations of the dihedral group
$\mathsf{D}_{2M}$ are instead the $2$ one-dimensional representations $U_0^\pm$ and the $(M-1)/2$ two-dimensional representations $R_k$, with $k=1,\ldots,(M-1)/2$. These representations are defined exactly as in \cref{eq:DihIrreps}.

Restriction from $\mathsf{O}(2)$ to $\mathsf{D}_{2M}$ acts on irreps in the following way:
\begin{subequations}
\label{eq:branching}
\begin{align}
    \left.\mathcal{U}_{0}^{\pm}\right|_{\mathsf{D}_{2M}} & = U_{0}^{\pm},
    \\
    \left.\mathcal{R}_{k}\right|_{\mathsf{D}_{2M}} & = R_{f_{M}\left(k\right)},
\end{align}
\end{subequations}
where we have introduced the function
\begin{equation}
\label{eq:branching2}
    f_{M}\left(k\right)\equiv\frac{\arccos\left[\cos\left(2k\pi/M\right)\right]}{2\pi/M}
    =
    M\,\left|\frac{k}{M}-\left\lfloor \frac{k}{M}+\frac{1}{2}\right\rfloor \right|
\end{equation}
and defined the (reducible) representations \begin{subequations}
\begin{align}
R_{0} & \equiv U_{0}^{+}\oplus U_{0}^{-},\\
R_{M/2} & \equiv U_{M/2}^{+}\oplus U_{M/2}^{-}.
\end{align}
\label{eq:branching3}
\end{subequations}

\begin{figure}
    \centering
    \includegraphics[width=.875\textwidth]{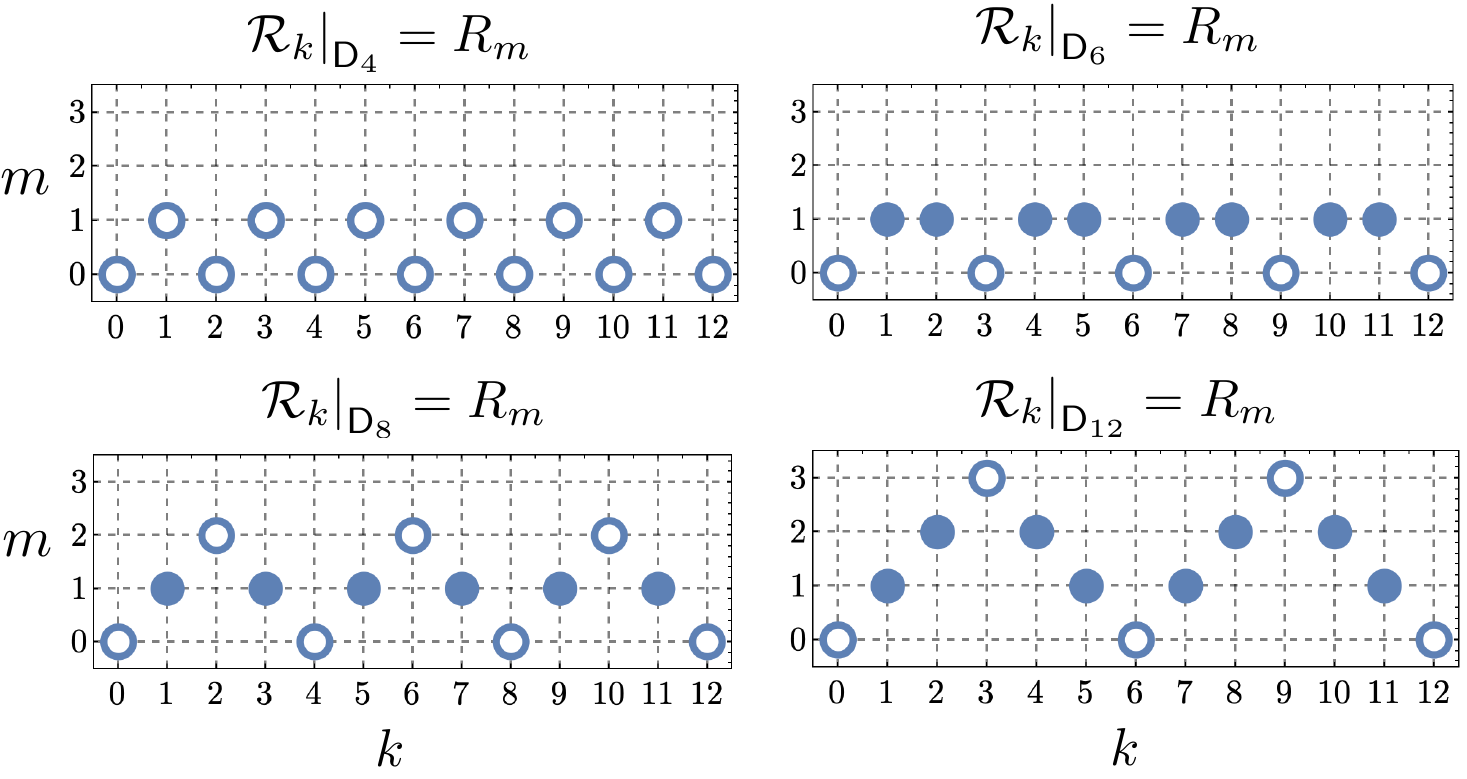}
    \caption{
    Visual illustration of the branching rules \labelcref{eq:branching,eq:branching2,eq:branching3} for the restriction of $\mathsf{O}(2)$ irreps $\mathcal{R}_k$
    to dihedral subgroups $\mathsf{D}_{2M}\leq \mathsf{O}(2)$, for $M=2,3,4,6$ (top left, top right, bottom left, bottom right, respectively). Open (closed) circles indicate that the $\mathsf{O}(2)$ representation is reducible (irreducible) upon restriction.
    }
    \label{fig:O2add}
\end{figure}
See \cref{fig:O2add} for an explicit illustration of the branching rules \labelcref{eq:branching,eq:branching2,eq:branching3} in the case of dihedral groups of low degree $M$.

\subsubsection{Tensor representations}

The orthogonal group $\mathsf{O}(2)$ has a natural action on real-valued, $2$-dimensional, rank-$n$ tensors of the form $T_{i_{1}\cdots i_{n}}$, given by
\begin{equation}
    T_{i_{1}\cdots i_{n}}\xrightarrow{g\in \mathsf{O}(2)}(g\cdot T)_{i_{1}\cdots i_{n}}
    \equiv
    \left(\prod_{k=1}^{n}\mathcal{R}_{1}(g)_{i_{k}j_{k}}\right)T_{j_{1}\cdots j_{n}},
    \label{eq:O2tensorAction}
\end{equation}
which may be thought of as ``rotating each index as a vector." We will be working exclusively in two dimensions, and so we define $\mathcal{T}_n = (\mathbb{R}^2)^{\otimes n}$ as the vector space of real $2$-dimensional, rank-$n$ tensors. It is clear then that $\mathsf{O}(2)$ acts \labelcref{eq:O2tensorAction} on $\mathcal{T}_n$ via the representation $\bigotimes_{k=1}^n \mathcal{R}_1$, which will reduce into a direct sum of irreducible $\mathsf{O}(2)$-representations according to the decomposition rules given in  \cref{eq:LRO2}.

Consider the vector space $\mathcal{T}_2$ of rank-2 tensors.  Tensors of this type are especially relevant for viscometry, since the strain tensor $s_{ij}\equiv\partial_i v_j$ is an element of this space. \cref{eq:LRO2} then tells us that  the action \labelcref{eq:O2tensorAction} of $\mathsf{O}(2)$ on $\mathcal{T}_2$ is reducible:
\begin{equation}
    \mathcal{R}_1 \otimes \mathcal{R}_1 = \mathcal{U}_0^+ \oplus \mathcal{U}_0^- \oplus \mathcal{R}_2 
    \label{eq:R1R1}.
\end{equation}
An explicit basis of $\mathcal{T}_2$ that achieves this block diagonalization is
\begin{equation}
    \left\{ \delta_{ij},\epsilon_{ij},\sigma^x_{ij},\sigma^z_{ij}\right\} \equiv\left\{ \left[\begin{array}{cc}
1 & 0\\
0 & 1
\end{array}\right]_{ij},\left[\begin{array}{cc}
0 & 1\\
-1 & 0
\end{array}\right]_{ij},\left[\begin{array}{cc}
0 & 1\\
1 & 0
\end{array}\right]_{ij},\left[\begin{array}{cc}
1 & 0\\
0 & -1
\end{array}\right]_{ij}\right\} 
\end{equation}
where, if $v\in\mathcal{A}$ is understood to mean that the vector
 $v\in\mathcal{T}_2$ lies in the subspace transforming exclusively under the
representation $\mathcal{A}$, we have that 
\begin{subequations}
\begin{align}
\delta_{ij} & \in\mathcal{U}_{0}^{+},\\
\epsilon_{ij} & \in\mathcal{U}_{0}^{-},\\
\left\{ \sigma^z_{ij},\sigma^x_{ij}\right\}  & \in\mathcal{R}_{2}.
\label{eq:muR2}
\end{align}
\end{subequations}
Illustrated explicitly for a given $T_{ij}\in\mathcal{T}_2$, we see that 
\begin{align}T_{ij} & =\left(\frac{\delta_{kl}}{\sqrt{2}}T_{kl}\right)\frac{\delta_{ij}}{\sqrt{2}}+\left(\frac{\epsilon_{kl}}{\sqrt{2}}T_{kl}\right)\frac{\epsilon_{ij}}{\sqrt{2}}+\left(\frac{\sigma_{kl}^{x}}{\sqrt{2}}T_{kl}\right)\frac{\sigma_{ij}^{x}}{\sqrt{2}}+\left(\frac{\sigma_{kl}^{z}}{\sqrt{2}}T_{kl}\right)\frac{\sigma_{ij}^{z}}{\sqrt{2}}\\
 & =\frac{1}{2}\left[\begin{array}{cc}
T_{xx}+T_{yy} & 0\\
0 & T_{xx}+T_{yy}
\end{array}\right]_{ij}+\frac{1}{2}\left[\begin{array}{cc}
0 & T_{xy}-T_{yx}\\
T_{yx}-T_{xy} & 0
\end{array}\right]_{ij}+\frac{1}{2}\left[\begin{array}{cc}
T_{xx}-T_{yy} & T_{xy}+T_{yx}\\
T_{xy}+T_{yx} & T_{yy}-T_{xx}
\end{array}\right]_{ij},
\end{align}
which is nothing other than the familiar statement that rank-2 tensors decompose into a trace, an antisymmetric, and a traceless symmetric ``part" (i.e. projection into an irreducible subspace), with this decomposition preserved under rotations and reflections.

Let us now restrict from the action \labelcref{eq:R1R1} of $\mathsf{O}(2)$ on $\mathcal{T}_2$ to the action of $\mathsf{D}_8$ on $\mathcal{T}_2$. Then we see from the branching rules \labelcref{eq:branching} that $\left.\mathcal{R}_{2}\right|_{\mathsf{D}_{8}}=U_{2}^{+}\oplus U_{2}^{-}$ and hence
\begin{equation}
    \left.\mathcal{R}_{1}\otimes\mathcal{R}_{1}\right|_{\mathsf{D}_{8}}=U_{0}^{+}\oplus U_{0}^{-}\oplus U_{2}^{+}\oplus U_{2}^{-}.
    \label{eq:T2D8}
\end{equation}
The reduction of $\mathcal{R}_2\to R_2 = U_2^+\oplus U^-_2$ to two $1$-dimensional irreducible representations upon restriction to $\mathsf{D}_8$ is precisely the mechanism responsible for the splitting of shear viscosity $\eta \to \eta_+, \eta_\times$ when rotational symmetry of the Fermi surface is broken in favor of square symmetry. This can be seen by the fact that the two viscosity tensor terms $\eta_{\times}\sigma_{ij}^{x}\sigma_{kl}^{x}$ and $\eta_{+}\sigma_{ij}^{z}\sigma_{jk}^{z}$ pick out rate of strain tensors that live in this symmetry sector. Similarly, the decomposition \labelcref{eq:T2D8} tells us that there will generically be viscosity tensor terms that pick out fluid motion living in the $U^+_0$ and $U^-_0$ irreducible representations: these are precisely the bulk viscosity $\zeta\delta_{ij}\delta_{kl}$ and rotational viscosity $\eta_{\circ}\epsilon_{ij}\epsilon_{kl}$, respectively.

Finally, consider the rank-4 tensor space $\mathcal{T}_4$, of which the viscosity tensor $\eta_{ijkl}$ is an element. \cref{eq:LRO2} then tells us that the action \labelcref{eq:O2tensorAction} on $\mathcal{T}_4$ is reducible as
\begin{equation}
    \otimes^{4}\mathcal{R}_{1}=3\mathcal{U}_{0}^{+}\oplus3\mathcal{U}_{0}^{-}\oplus4\mathcal{R}_{2}\oplus\mathcal{R}_{4}.
    \label{eq:T4action}
\end{equation}
An explicit basis of $\mathcal{T}_4$ that achieves this block diagonalization is given by
\begin{equation}
\begin{array}{ccc}
\delta\delta\in\mathcal{U}_{0}^{+}&\epsilon\epsilon\in\mathcal{U}_{0}^{+}&\left(\sigma^{x}\sigma^{x}+\sigma^{z}\sigma^{z}\right)\in\mathcal{U}_{0}^{+}\\\left(\delta\epsilon+\epsilon\delta\right)\in\mathcal{U}_{0}^{-}&\left(\delta\epsilon-\epsilon\delta\right)\in\widehat{\mathcal{U}}_{0}^{-}&\left(\sigma^{x}\sigma^{z}-\sigma^{z}\sigma^{x}\right)\in\widehat{\mathcal{U}}_{0}^{-}\\\left\{ \left(\delta\sigma^{x}+\sigma^{x}\delta\right),\left(\delta\sigma^{z}+\sigma^{z}\delta\right)\right\} \in\mathcal{R}_{2}&\left\{ \left(\epsilon\sigma^{x}+\sigma^{x}\epsilon\right),\left(\epsilon\sigma^{z}+\sigma^{z}\epsilon\right)\right\} \in\mathcal{R}_{2}&\left\{ \left(\delta\sigma^{x}-\sigma^{x}\delta\right),\left(\delta\sigma^{z}-\sigma^{z}\delta\right)\right\} \in\widehat{\mathcal{R}}_{2}\\\left\{ \left(\epsilon\sigma^{x}-\sigma^{x}\epsilon\right),\left(\epsilon\sigma^{z}-\sigma^{z}\epsilon\right)\right\} \in\widehat{\mathcal{R}}_{2}&\left\{ \left(\sigma^{x}\sigma^{x}-\sigma^{z}\sigma^{z}\right),\left(\sigma^{x}\sigma^{z}+\sigma^{z}\sigma^{x}\right)\right\} \in\mathcal{R}_{4}&
\end{array}
\label{eq:T4basis}
\end{equation}
In \cref{eq:T4basis}, we have omitted $i,j,k,l$ indices, with their placement implied by the order of tensors; the $i,j$ indices go on the first (left) tensor in any product, and the $k,l$ indices on the second (right) tensor. For example, $\delta\delta=\delta_{ij}\delta_{kl}$. We have also further diagonalized equivalent $\mathsf{O}(2)$ irreps according to their parity under time reversal ($ij\leftrightarrow kl$, or equivalently in the above notation, switching the order of tensors in any product), with extra hats being put on $\mathsf{O}(2)$ irreps that are time-reversal odd.

To say that a viscosity tensor $\eta_{ijkl}$ is isotropic, i.e. $\mathsf{O}(2)$-invariant, is simply the statement that $\eta_{ijkl}\in\mathcal{U}_0^+$, i.e. it transforms trivially under the action \labelcref{eq:O2tensorAction}. From the basis \labelcref{eq:T4basis}, we can already see the generality of the isotropic ($M=\infty$) viscosity tensor \labelcref{eq:viscTensor} from the main text; only the terms that belong to the trivial representation $\mathcal{U}_0^+$ may appear in the isotropic viscosity tensor \labelcref{eq:viscTensor}. For the isotropic tensor \labelcref{eq:viscTensor}, we have excluded the $\epsilon_{ij}\epsilon_{kl}$ tensor despite it belonging to the trivial representation $\mathcal{U}_0^+$, simply because the corresponding component is proportional to the antisymmetric part $\epsilon_{ij}T_{ij}$ of a stress tensor $T_{ij}$, which much always vanish  by angular momentum conservation in an isotropic theory.

If we relax our notion of isotropy and no longer demand invariance under reflection, then tensors belonging to the $\mathcal{U}_0^-$ representation may also be included, i.e. the first \emph{six} tensors in \cref{eq:T4basis}. These six tensors exactly match those found in recent work \cite{viscTensors2020} enumerating the most general viscosity tensors allowed in an ``isotropic" (allowing for non-trivial reflection parity) fluid. Furthermore, the tensors given in \cref{eq:T4basis} also contain those found in recent work \cite{hallViscPRX2020} on Hall viscosities in anisotropic fluids with broken time-reversal symmetry.

Upon restriction from $\mathsf{O}(2)$ to $\mathsf{D}_{12}$, we see from the branching rules \labelcref{eq:branching} that none of the non-trivial $\mathcal{T}_4$ basis elements (not lying in the irrep $\mathcal{U}_0^+$) \labelcref{eq:T4basis} become trivial, i.e. we get no new invariant tensors upon restriction to $\mathsf{D}_{12}$. However, in this case, anisotropy allows the stress tensor $T_{ij}$ to have an antisymmetric component $\epsilon_{ij}T_{ij}\neq 0$, and so we now include the invariant tensor $\epsilon_{ij}\epsilon_{kl}$ in \cref{eq:viscTensor}.

Upon restriction from $\mathsf{O}(2)$ to $\mathsf{D}_{8}$, the $\mathcal{R}_4$ irrep decomposes and contains a trivial $\mathsf{D}_8$ irrep, since $\left.\mathcal{R}_{4}\right|_{\mathsf{D}_{8}}=U_{0}^{+}\oplus U_{0}^{-}$ per \cref{eq:branching}. This means that we may now use the first basis element in the $\mathcal{R}_4$ subspace \labelcref{eq:T4basis} in our $\mathsf{D}_8$-invariant viscosity tensor. This vector $(\sigma_{ij}^{x}\sigma_{kl}^{x}-\sigma_{ij}^{z}\sigma_{kl}^{z})$, when taken into linear combinations with the always-trivial vector $(\sigma_{ij}^{x}\sigma_{kl}^{x}+\sigma_{ij}^{z}\sigma_{kl}^{z})$, then allows the tensors $\sigma^x_{ij}\sigma^x_{kl}$ and $\sigma^z_{ij}\sigma^z_{kl}$ to appear \emph{independently} in the $\mathsf{D}_8$ viscosity tensor \labelcref{eq:viscTensor}. This is again the origin of the shear viscosity splitting $\eta\to\eta_+,\eta_\times$ upon restriction to $\mathsf{D}_8$.

\begin{table*}[t!]
\includegraphics[width=\textwidth]{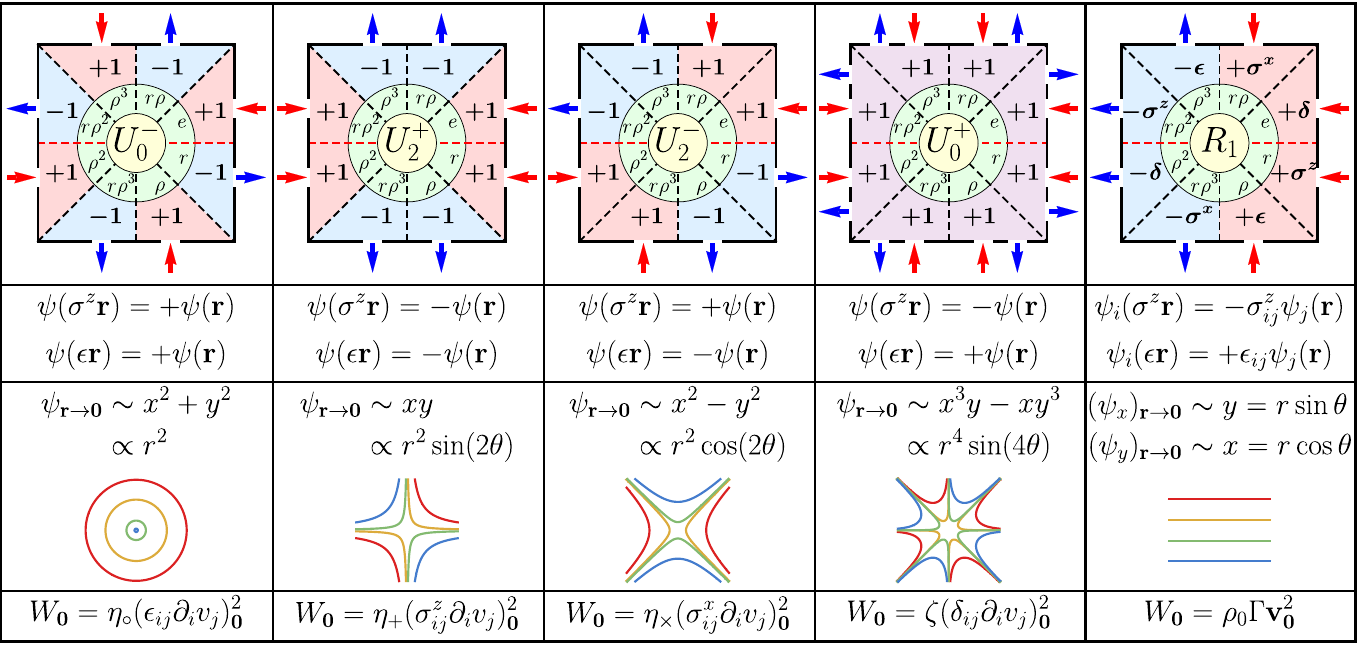}
\caption[]{
\textit{First row}: Visualizations of the dihedral group $\mathsf{D}_8$ (green) and its five irreducible representations (yellow), alongside viscometer boundary conditions (red/blue arrows) of matching symmetry. \textit{Second row}: Reflection and rotational implications of $\mathsf{D}_8$-irreducible boundary conditions. The stream function is defined via a right-handed cross product, which gives $\psi\hat{\mathbf{z}}$ (pseudovector) an extra sign change under reflections relative to the current (vector) boundary conditions. For the 2-dimensional irrep $R_1$, the solution $\psi$ is a linear combination of two functions $\psi_{x,y}$, which transform among each other under the action of $\mathsf{D}_8$. \textit{Third row}: Symmetry constrains the functional form of the fluid flow (i.e. streamlines) near a high symmetry point, the square center. \textit{Fourth row}: Symmetry-constrained flow at the square center $\mathbf{r}=\mathbf{0}$ restricts the center heat $W_\mathbf{0}$ to only the dissipative coefficient in the matching symmetry sector of $\mathsf{D}_8$.
}
\label{table:irrepsFULL}
\end{table*}

See \cref{table:irrepsFULL} for a visualization of the five $\mathsf{D}_8$ irreps realized as boundary conditions on our dihedral viscometer. By enforcing current boundary conditions with symmetry of a selected $\mathsf{D}_8$ irrep, one may restrict dissipation at the square center to selected dissipative coefficients (e.g. viscosity components discussed above) as desired.

Finally, similar considerations for the restriction of $\mathsf{O}(2)$ to $\mathsf{D}_4$ give the following symmetry-allowed viscosity tensor in $\mathsf{D}_4$ fluids:
\begin{align}
\label{eq:D4tensor}\eta_{ijkl}&=\zeta\left(\delta_{ij}\delta_{kl}\right)+\eta_{\circ}\left(\epsilon_{ij}\epsilon_{kl}\right)+\eta_{\times}\left(\sigma_{ij}^{x}\sigma_{kl}^{x}\right)+\eta_{+}\left(\sigma_{ij}^{z}\sigma_{kl}^{z}\right)\nonumber\\&\qquad+\eta_{\zeta+}\left(\delta_{ij}\sigma_{kl}^{z}+\sigma_{ij}^{z}\delta_{kl}\right)+\eta_{\circ\times}\left(\epsilon_{ij}\sigma_{kl}^{x}+\sigma{}_{ij}^{x}\epsilon_{kl}\right)\\&\qquad+\widehat{\eta}_{\zeta+}\left(\delta_{ij}\sigma_{kl}^{z}-\sigma_{ij}^{z}\delta_{kl}\right)+\widehat{\eta}_{\circ\times}\left(\epsilon_{ij}\sigma_{kl}^{x}-\sigma{}_{ij}^{x}\epsilon_{kl}\right)\nonumber
\end{align}
Taken in addition to \cref{eq:viscTensor}, \cref{eq:D4tensor} completes the specification of the most general viscosity tensor allowed in dihedral fluids \emph{of any degree M}. In \cref{eq:D4tensor}, we have used hats to indicate the $\mathsf{D}_4$ viscosities which are time-reversal odd; these viscosities will only appear in  $\mathsf{D}_4$ fluids which, in addition to their low rotational symmetry, have also broken time-reversal symmetry.

\section{Comparison of our viscometry and channel flow techniques}

In this section, we compare our viscometry technique to those based on flow profiles in long channels, a more conventional probe of electronic viscosity. In long, one-dimensional channels with no-slip boundary conditions at the walls, viscous  flow leads to a parabolic (Poiseuille) velocity profile \cite{lucasreview17}. The curvature of this parabolic velocity profile is set by (a component of) the fluid viscosity, with larger viscosities giving rise to smaller curvature and vice-versa; by measuring this velocity profile curvature (or the integrated flow it induces), the relevant viscosity component may be inferred.

As it is the most general possible case, we consider the hydrodynamic flow of a $\mathsf{D}_4$-invariant fluid \labelcref{eq:D4tensor}, forced by an applied field (i.e. pressure gradient) through an infinite 1D channel $(X,Y)\in\mathbb{R}\times[-W/2,W/2]$ of transverse width $W$. Channel flows of $\mathsf{D}_4$-invariant Dirac fluids (e.g. charge neutral graphene) were studied in \cite{link}, in which a viscometry procedure was also proposed. Their proposal involved measuring the curvature of the resulting Poiseuille channel profile, \emph{as a function of the relative angle between the channel and fluid's symmetry/crystallographic axes}. The suggested procedure then exploits this angular freedom to (in principle) extract multiple viscosity components of the $\mathsf{D}_4$ fluid.

We assume that the channel coordinates $(X,Y)$ are rotated
\begin{align}
        \left[\begin{array}{c} X \\ Y \end{array}\right] =\left[\begin{array}{cc} \cos\theta &\ \sin\theta \\ -\sin\theta &\ \cos\theta \end{array}\right] \left[\begin{array}{c} x \\ y \end{array}\right]
\end{align}
by an angle $\theta$ relative to the $\mathsf{D}_4$ fluid coordinates $(x,y)$. As described above, the fluid is forced through the channel by an electric field of strength $E_X$, applied in the positive $X$-direction. Assuming no-slip $v_X = 0$ at the channel walls $|Y|=W/2$, the static velocity profile is then the parabolic, Poiseulle solution \cite{lucasreview17} 
\begin{equation}
    v_X(Y) = \frac{n e  E_X}{2\eta_{XYXY}(\theta)} \left(\frac{W^2}{4}-Y^2\right)
\end{equation}
where $\eta_{XYXY}(\theta)$ is the relevant channel viscosity component, properly rotated from the fluid coordinates $(x,y)$ via \cref{eq:O2tensorAction}; using \cref{eq:O2tensorAction,eq:D4tensor}, this component is computed to be
\begin{equation}
    \eta_{XYXY}\left(\theta\right)
    =
    \frac{1}{2}\left(2\eta_{\circ}+\eta_{\times}+\eta_{+}\right)+\left(4\eta_{\circ\times}\right)\cos\left(2\theta\right)+\left(\eta_{\times}-\eta_{+}\right)\cos\left(4\theta\right),
\label{eq:rotatedD4}
\end{equation}
or, equivalently, in Cartesian coordinates of the fluid:
\begin{equation}
    \eta_{XYXY}\left(\theta\right)=\left(\eta_{xxxx}-\eta_{xxyy}-\eta_{xyyx}-\eta_{yxyx}-\eta_{yyxx}+\eta_{yyyy}\right)\cos^{2}\theta\sin^{2}\theta\\+\left(\eta_{xyxy}\right)\cos^{4}\theta+\left(\eta_{yxyx}\right)\sin^{4}\theta.
\label{eq:rotatedD4cart}
\end{equation}
Our approach possesses several manifest advantages over such rotated channel flow experiments.

Firstly, as can be seen from both \cref{eq:rotatedD4} and \cref{eq:rotatedD4cart}, such rotated Poiseuille flows can distinguish at most $3$ unique viscosity components, of the $8$ total \labelcref{eq:D4tensor} allowed in $\mathsf{D}_4$ fluids ($6$ total if time-reversal is a symmetry). By contrast, we expect the $4$ boundary condition irreps in $\mathsf{D}_4$ (i.e. $U^{0,1}_\pm$) to give $4$ distinct heat measurements at the center of a square/rectangle viscometer, from which (linear combinations of) $4$ of the $\mathsf{D}_4$ viscosity components \labelcref{eq:D4tensor} may be inferred.  Moreover, for higher symmetry cases (to which our approach naturally generalizes), it is clear that our viscometry will continue to distinguish strictly more viscosities than rotated channels (e.g. $\eta_\circ$ in $\mathsf{D}_8$ fluids).

Secondly, even in fluids of exceptionally-low $\mathsf{D}_4$ symmetry, for which irreducible boundary conditions are not enough to uniquely isolate all viscosities, our viscometry nevertheless continues to group viscosities according to their symmetry \emph{class}. For example, $U_0^-$ boundary currents on a square sample of $\mathsf{D}_4$ fluid would lead to center heating from $\eta_\circ$, $\eta_\times$, $\eta_{\circ\times}$ (and thus be unable to distinguish them) --- but those $3$ alone, and none of the other $5$ allowed in $\mathsf{D}_4$. As a caveat: in order to use our framework to measure viscosities in a $\mathsf{D}_4$-invariant fluid, one will need to compare experimental heating measurements with e.g. detailed hydrodynamic simulations.

Thirdly, the feasibility of such rotated-channel experiments relies on the ability to cut the requisite channel samples at various angles relative to the crystal axes. In order to distinguish even the $3$ channel viscosities \labelcref{eq:rotatedD4} just discussed, $3$ different channel angles must be used, therefore requiring at least one mesoscopic sample misaligned with the crystallographic axes. By contrast, our viscometry relies not on the \emph{geometry} of the boundary, but rather its \emph{symmetry}. For example, for the $\mathsf{D}_8$ fluids discussed in the main text, square samples/boundaries are not required; isolated centered heating will still be guaranteed even with circular samples/boundaries, so long as the current boundary conditions remain $\mathsf{D}_8$-irreducible.



\section{Kinetic theory}
\label{app:kinetic}

In this appendix, we discuss the extent to which our argument in the main text generalizes to account for ballistic effects.  For a sufficiently weakly interacting electron fluid, we can solve Boltzmann equations to calculate transport coefficients beyond the hydrodynamic regime \cite{hartnoll1705}.  As in the main text, we study time-independent flows within linear response.  Letting $\varphi(x,p) = f(x,p) - f_{\mathrm{eq}}(x,p)$ denote the deviation of the distribution function of kinetic theory away from equilibrium, the form of the kinetic equations is schematically: \begin{equation}
    v\left(p\right)\cdot\partial_{x}\left|\varphi(x)\right\rangle +W\left|\varphi(x)\right\rangle =0,
    \label{eq:boltzmann}
\end{equation}
where $v(p) = \partial_p \epsilon(p)$ denotes the microscopic (single-particle) group velocity arising from the single-particle dispersion relation, and $W$ denotes the linearized collision integral. We have also introduced Dirac notation to emphasize that the function $\varphi(x,p)$ is to be regarded as an infinite-dimensional vector in momentum space, so that \begin{equation}
    W\left|\varphi(x)\right\rangle =\int\mathrm{d}p^{\prime}\;W\left(p,p^{\prime}\right)\varphi\left(x,p^{\prime}\right).
    \label{eq:linCollision}
\end{equation}
We assume, as usual, that the collision integral is local in space. 

Without specifying any microscopic details, what can we say on the basis of symmetry alone?   As in the main text, let us imagine solving this Boltzmann equation \labelcref{eq:boltzmann} in a region $\Sigma$, which admits a natural group action by a symmetry group $G$, by which we mean the spatial geometry \emph{and} the dispersion relation are $G$-invariant. Now suppose the spatial geometry contains a point $x^\star\in\Sigma$ which is fixed by the action of $G$, i.e. $g\cdot x^\star = x^\star$ for all $g\in G$. Consider a solution $\varphi^\star(p)\equiv\varphi(x=x^{\star},p)$ of the Boltzmann equation \labelcref{eq:boltzmann}, evaluated at this fixed point. Then the action of $G$ on the vector space of fixed-point-evaluated distributions $V=\{|\varphi^\star\rangle\}$, given by
\begin{equation}
     g\cdot \varphi^\star(p) \equiv \varphi(g^{-1}\cdot x^\star, g^{-1}\cdot p) = \varphi(x^\star, g^{-1}\cdot p)=\varphi^\star(g^{-1}\cdot p),
     \label{eq:action}
\end{equation}
restricts to only the momentum-dependence.

Since $G$ is assumed to be a group of symmetries, we know that the linearized collision integral $W$ \labelcref{eq:linCollision} is invariant under the group action \labelcref{eq:action}. But then Schur's lemma \cite{tung} tells us that the vector space $V$ of possible fixed point momentum distributions $|\varphi^\star\rangle$'s may be written as a direct sum $V=\bigoplus_R\bigoplus_n V_{R;n}$ of $G$-irreducible subspaces $V_{R;n}$, each acted upon by the action \labelcref{eq:action} of $G$ according to an irrep $R$ of $G$, so that $W$ acts proportionally to the identity on each irreducible subspace $V_{R;n}$. This allows us to write
\begin{equation}
    W = \sum_{R} \sum_{n} w_{R;n} P_{R;n},
    \label{eq:irrepDecomp}
\end{equation}
where $P_{R;n}$ denotes a projector onto $V_{R;n}$, and $w_{R;n}$ are the proportionality constants. We have introduced the extra label $n$ to account for the inevitable appearance of multiple copies of each irrep $R$; it is entirely analogous to the quantum number $n$ that appears in the  wave functions $\psi_{nlm}$ of a rotationally-invariant quantum mechanical model, where only $lm$ indices specify the rotational symmetry. 

Note that, by the decomposition \labelcref{eq:irrepDecomp}, the irreducible subspaces $V_{R;n}$ are also eigenspaces of the linearized collision integral $W$, with the corresponding eigenvalues $w_{R;n}$ playing the same role mathematically as the viscosity components described in the main text. In the context of kinetic theory, the collision integral eigenvalues $w_{R;n}$ have the physical interpretation as \emph{scattering rates} associated with various scattering mechanisms/pathways.

If we now choose boundary conditions which transform exclusively under a given irrep $R^\prime$ of the symmetry group $G$, then the function $\varphi^\star(p)$, as the solution of a $G$-invariant differential equation \labelcref{eq:boltzmann} with $R^\prime$-covariant boundary conditions, must also transform according to the irrep $R^\prime$ under the group action \labelcref{eq:action}. In other words, $\left|\varphi^\star\right\rangle \in\bigoplus_{n}V_{R^\prime;n}$. This result then allows us to express the vector $\left|\varphi^\star\right\rangle\equiv\left|\varphi^\star_{R^\prime}\right\rangle$ as
\begin{equation}
    \left|\varphi_{R^{\prime}}^{\star}\right\rangle =\sum_{n}\left\langle \varphi_{R^{\prime};n}^{\star}\left|\varphi_{R^{\prime}}^{\star}\right.\right\rangle \left|\varphi_{R^{\prime};n}^{\star}\right\rangle 
\end{equation}
where $|\varphi_{R^{\prime};n}^{\star}\rangle\in V_{R^\prime;n}$ for each $n$.

The fixed point heating $Q(x^\star)$ is then calculated in kinetic theory as
\begin{equation}
    Q\left(x^{\star}\right)=\langle\varphi_{R^{\prime}}^{\star}|\,W\,|\varphi_{R^{\prime}}^{\star}\rangle=\sum_{n}w_{R^{\prime};n}\left|\left\langle \varphi_{R^{\prime};n}^{\star}\left|\varphi_{R^{\prime}}^{\star}\right.\right\rangle \right|^{2}.
    \label{eq:ktHeat}
\end{equation}
Importantly, the scattering rates that contribute to the fixed point heat $Q(x^\star)$ \labelcref{eq:ktHeat} are isolated to only those $w_{R;n}$ in \cref{eq:irrepDecomp} for which $R=R^\prime$, the irrep specified by the boundary conditions. We therefore conclude: only dissipative mechanisms that couple to momentum functions $\varphi^\star(p)$ belonging to the same irrep $R^\prime$ as the boundary conditions contribute to heat at a fixed point $x=x^\star$.  In the hydrodynamic regime, these dissipative mechanisms are viscosities (to leading order in the small parameter $\ell_{\mathrm{ee}}/L$, with $L$ the characteristic length scale of $\Sigma$). The fixed point heat \labelcref{eq:ktHeat} is analogous to the selected isolation of a single term in the $\mathsf{D}_8$ heating decomposition \labelcref{eq:velocityHeat} given in the main text (though in that case, there are no repeated irreps, so there is no $n$ index).  


Finally, we address a subtlety that arises when we instead allow the boundary condition symmetry group $H$ to be a \emph{subgroup} of the fluid symmetry group $G$, in which case we must slightly generalize \cref{eq:ktHeat}. For concreteness, let us now take boundary conditions which transform under a given irrep $S^\prime$ of $H$. When the irreps $R$ of $G$ are restricted to $H$, they generate representations $R|_H$ of $H$, which are in general reducible with respect to $H$.   So in this case, fixed point heating can be generated by all irreps $R$ of $G$ for which the irreducible decomposition of $R|_{H}$ contains $S^\prime$, the boundary condition irrep of $H$.   Put another way, the smaller symmetry group $H$ of the device determines the constrained heating, not the larger fluid symmetry group $G$.

As a result, if an isotropic $G=\mathsf{O}(2)$ fluid is placed in a viscometer with $S^\prime=U_0^-$ boundary conditions, irreducible with respect to a dihedral subgroup such as $H=\mathsf{D}_8$, then there will be extremely small heating at a fixed point in the hydrodynamic regime.  The leading contribution to heat generated at the center of the device comes from kinetic theory modes $|\varphi_n\rangle$ that are in the $\mathcal{R}_4$ representation of $G=\mathsf{O}(2)$, since the decomposition $\left.\mathcal{R}_{4}\right|_{\mathsf{D}_{8}}=U_{0}^{+}\oplus U_{0}^{-}$ contains $S^\prime$.   In the hydrodynamic regime, one finds that in a device of size $w$, with electron-electron scattering length $\ell_{\mathrm{ee}}$, \cite{lucas1612} \begin{subequations}
\begin{align}
    w_{\mathcal{R}_4;n} &\sim \frac{1}{\ell_{\mathrm{ee}}}, \\
    \langle \varphi_{\mathcal{R}_4;n} | \varphi_{U_0^-} \rangle &\sim \left(\frac{\ell_{\mathrm{ee}}}{w}\right)^3 v_{\mathrm{typ}} \sim \left(\frac{\ell_{\mathrm{ee}}}{w}\right)^3 \frac{I_0}{w},
\end{align}
\label{eq:corrections}
\end{subequations}
where $I_0$ is the total current that enters/exits through one of the contacts \footnote{The scaling $v_{\mathrm{typ}} \sim I_0/w$ follows from dimensional analysis, as $I_0$ is (up to overall prefactors that are not relevant here) the integral over the one-dimensional boundary of velocity.}. We therefore conclude that (in the limit $a/w\rightarrow 0$, so that dimensional analysis can be trusted) the total fixed point heating obeys \begin{equation}
    Q(x^\star) \sim \frac{1}{\ell_{\mathrm{ee}}} \left[\left(\frac{\ell_{\mathrm{ee}}}{w}\right)^3 \frac{I_0}{w}\right]^2 \sim \frac{\ell_{\mathrm{ee}}^5 I_0^2}{w^8}.
    \label{eq:ballCorrection}
\end{equation}
In a Fermi liquid where $\ell_{\mathrm{ee}} \sim T^{-2}$, the heating $Q(x^\star) \sim \ell_{\mathrm{ee}}^5 w^{-8}$ is extremely small;  the $T$ and $w$  dependence of $Q(x^\star)$ is extreme and remains a diagnostic for the absence of rotational viscosity in such a system.  After all, the rotational heating (in the hydrodynamic regime) instead scales as 
\begin{equation}
    W_{\circ}=\eta_{\circ}\left(\epsilon_{ij}\partial_{i}v_{j}\right)^{2}\sim \ell_{\mathrm{ee}} \left(\frac{1}{w}\frac{I_0}{w}\right)^2 \sim \frac{\ell_{\mathrm{ee}}I_0^2}{w^4},
\end{equation}
which is easily distinguishable.

Boundary conditions with full $\mathcal{U}^-_0$ symmetry correspond to the Taylor-Couette device geometry, i.e. constant, perfectly tangential $\mathbf{v}=v_\theta \hat{\theta}$ velocity everywhere along a circular/cylindrical boundary. Even though these boundary conditions satisfy $\langle \varphi_{\mathcal{R}_k;n} | \varphi_{\mathcal{U}_0^-} \rangle=0$ and therefore set \emph{all} perturbative ballistic corrections \labelcref{eq:ballCorrection} to zero at the fixed point (i.e. the center of the circular geometry), they are physically unrealizable in an electronic system, for which only orthogonal currents can be readily controlled.




\section{Estimating temperature signal due to viscous heating}
\label{app:temp}

In this appendix, we give an order of magnitude estimate for the expected temperature variation $(\nabla^2 T)_\mathbf{0}$ to be measured at the center of the dihedral viscometer, described in the main text and reported in \cref{fig:heatTable}. In doing so, we consult recent experimental data for hydrodynamic electrons in doped monolayer graphene \cite{bandurin,kumar2017superballistic}; these works report the following parameter values appropriate for the onset of hydrodynamic behavior in monolayer graphene:
\begin{subequations}\begin{align}
    n & \approx 10^{12}\text{ cm}^{-2},\\
    T_{\text{e}} & \approx 100\text{ K}, \\
    \lambda & \approx 1\text{ }\mu\text{m},\\
    \nu & \approx 0.1 \text{ m/s}^2, \\
    \sigma & \approx 0.03\text{ siemens},
\end{align}\end{subequations}
where $n$ is the electron density, $T_\text{e}$ the electron temperature, $\lambda$ the Gurzhi length, $\nu$ the kinematic (shear) viscosity, and $\sigma$ the DC electrical conductivity. 

We begin by estimating the electronic thermal conductivity $\kappa$, which can -- within an order of magnitude, in current experimental devices -- be related to the electrical conductivity $\sigma$ via the Wiedemann-Franz relation
\begin{equation}
    \mathcal{L}\equiv\frac{\kappa}{\sigma T_{\text{e}}}\simeq\mathcal{L}_{0}=\frac{\pi^{2}}{3}\left(\frac{k_{\text{B}}}{e}\right)^{2}.
\end{equation}
Solving for $\kappa$ and substituting in monolayer graphene paramter values, we estimate
\begin{equation}
    \kappa\approx\frac{\pi^{2}}{3}\left(\frac{k_{\text{B}}}{e}\right)^{2}\sigma T_{\text{e}}\approx7.33\cdot10^{-8}\text{ W/K}.
\end{equation}
We will also require the shear viscosity $\eta=m n \nu$, where $m$ is the quasiparticle mass. In monolayer graphene we may estimate $m$ by equating the linear $mv_\text{F}$ and cyclotronic $\hbar k_\text{F}$ momenta, with Fermi wavevector $k_\text{F}=\sqrt{\pi n}$ in two dimensions and typical Fermi velocity  $v_\text{F}\approx 10^6 \text{ m/s}$ \cite{lucasreview17} in monolayer graphene. Altogether, this gives a shear viscosity
\begin{equation}
    \eta = \left(\frac{\hbar\sqrt{\pi n}}{v_{\text{F}}}\right)n\nu \approx1.87\cdot10^{-17}\text{ kg/s}.
\end{equation}
As anisotropic viscosity components (such as $\eta_+, \eta_\times$) have never been measured in experiment, we will further assume that all viscosity components $\eta_\alpha\approx \eta$ are approximately equal to the above shear viscosity in graphene.  For anisotropic electron hydrodynamics in ABA trilayer graphene, this assumption is justified by recent theoretical calculations in a microscropic model \cite{marvin2021}.

Finally, we apply dimensional analysis to restore units to the center heats $W_\mathbf{0}=\eta_\alpha (\partial v_\alpha)_\mathbf{0}$, and hence the center temperature variations $-(\nabla^2 T)_\mathbf{0}=W_\mathbf{0}/\kappa$ \labelcref{eq:poissonW}, numerically obtained from the dimensionless biharmonic equation \labelcref{eq:bi2} given in the main text. The magnitude of strain rates  appearing in the sample will depend on choice of experimental parameters $w$ (the size of the viscometer) and $I_0$ (the currents applied to the sample); we take
\begin{subequations}\begin{align}
    w & \approx 1\text{ }\mu\text{m}\\
    I_0 & \approx 100\text{ }\mu\text{A},
\end{align}\end{subequations}
where $w\approx\lambda$ is chosen so that ohmic effects do not dominate the onset of hydrodynamics, and $I_0$ is a current value typical for linear response experiments in such systems. Letting $(\overline{\partial v})_\mathbf{0}$ represent the dimensionless center strain rates obtained from \cref{eq:bi2}, we find
\begin{equation}
    -\left(\nabla^{2}T\right)_{\mathbf{0}} =\frac{\eta}{\kappa}\left[\frac{1}{w}\left(\frac{I_{0}}{new}\right)\right]^{2}\left(\overline{\partial v}\right)_{\mathbf{0}}^{2}\notag 
    \approx\left(1\text{ }\text{K}/\mu\text{m}^2\right)\left(\overline{\partial v}\right)_{\mathbf{0}}^{2}.
\label{eq:Testimate}
\end{equation}
Signals of this magnitude are easily detectable using existing local thermometry based on nitrogen-vacancy centers in diamond: see e.g. \cite{diamondTempSensing}.


\section{Advantages for experiments}
\label{app:experiments}

\begin{figure}
    \centering
    \includegraphics[width=\textwidth]{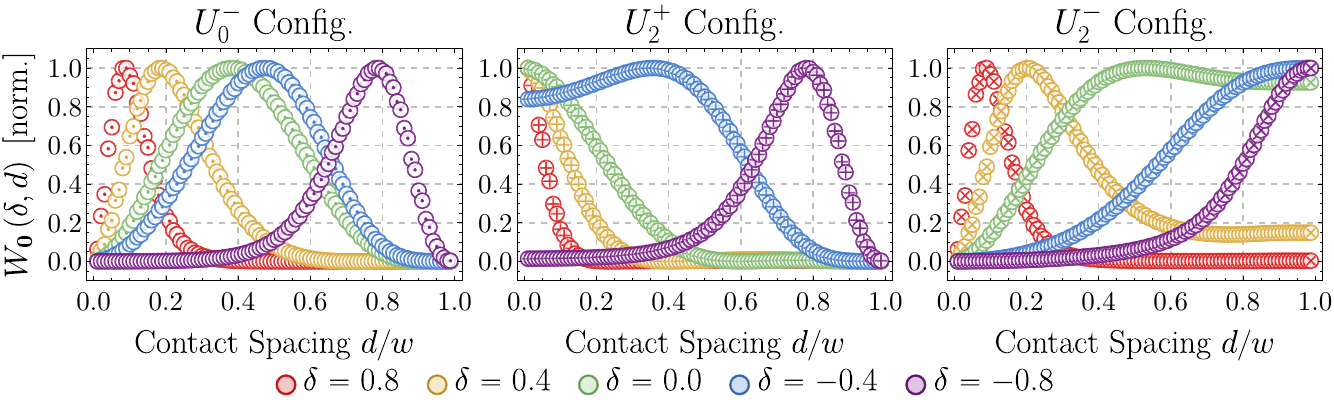}
    \caption{
    Reproduction of \cref{fig:invertHeats}, except now we have taken a relatively small Gurzhi length $\lambda/w=1/5$ (very strong ohmic scattering over the scale of the viscometer), as opposed to $\lambda/w=\infty$ (no ohmic scattering) in \cref{fig:invertHeats}. As this plot is nearly identical to \cref{fig:invertHeats}, we conclude that the shapes of these curves are extraordinarily insensitive to momentum-relaxing processes in an electron fluid.
    }
    \label{fig:invertHeatsOHM}
\end{figure}

\begin{figure}
    \centering
    \includegraphics[width=.55\textwidth]{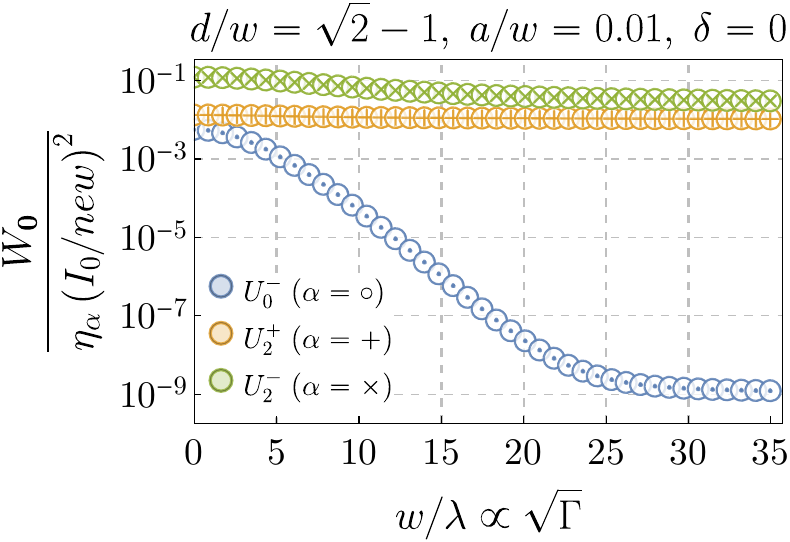}
    \caption{
    Center heat signal strength as a function of finite Gurzhi length $\lambda$, i.e. non-zero ohmic scattering rate $\Gamma$, for various $\mathsf{D}_8$-irreducible boundary conditions. A would-be electron fluid in an experiment of length-scale $w$ can only be typically regarded as a fluid, with momentum conserved to a good approximation, for at most $w/\lambda \lesssim 5$. Thus, the center heat signal is extremely insensitive to momentum-relaxing scattering, as long as we are still in the hydrodynamic regime.
    }
    \label{fig:ohmSignal}
\end{figure}

\begin{figure}
    \centering
    \includegraphics[width=.5\textwidth]{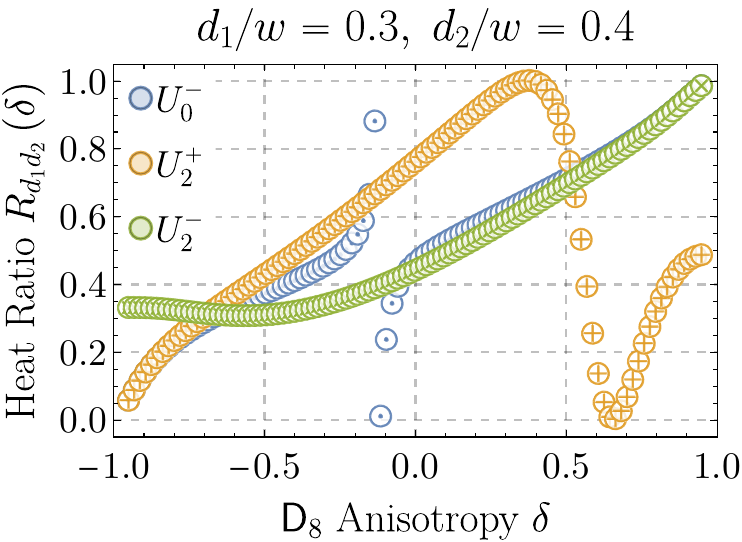}
    \caption{
    Plot of the ratio of heats $R_{d_{1}d_{2}}\left(\delta\right)\equiv W_{\mathbf{0},\delta}\left(d_{1}\right)/\left[W_{\mathbf{0},\delta}\left(d_{1}\right)+W_{\mathbf{0},\delta}\left(d_{2}\right)\right]$ at two different contact spacing values $d_1$ and $d_2$,  with $a/w=0.01$ and $\lambda/w=\infty$. Note that the yellow and green curves never fail the horizontal line test at the same pair of anisotropy values $\delta_a,\delta_b\in(-1,1)$. This implies that the 2 experimentally-determined heat ratios $\left(R_{d_{1}d_{2}}\right)_{U_{2}^{+}}$ and $\left(R_{d_{1}d_{2}}\right)_{U_{2}^{-}}$, constituting 4 total center heat measurements, are sufficient to uniquely determine $\delta$. The singular behavior of $\left(R_{d_{1}d_{2}}\right)_{U_{0}^{-}}$ near $\delta\approx -0.12$ corresponds to the closing and re-opening of the central $U_0^-$ vortex around that anisotropy value for $d_1/w=0.3$ (see \cref{fig:vortexDestruct,fig:rotDelta}).
    }
    \label{fig:d2d}
\end{figure}

\begin{figure}
    \centering
    \includegraphics[width=.95\textwidth]{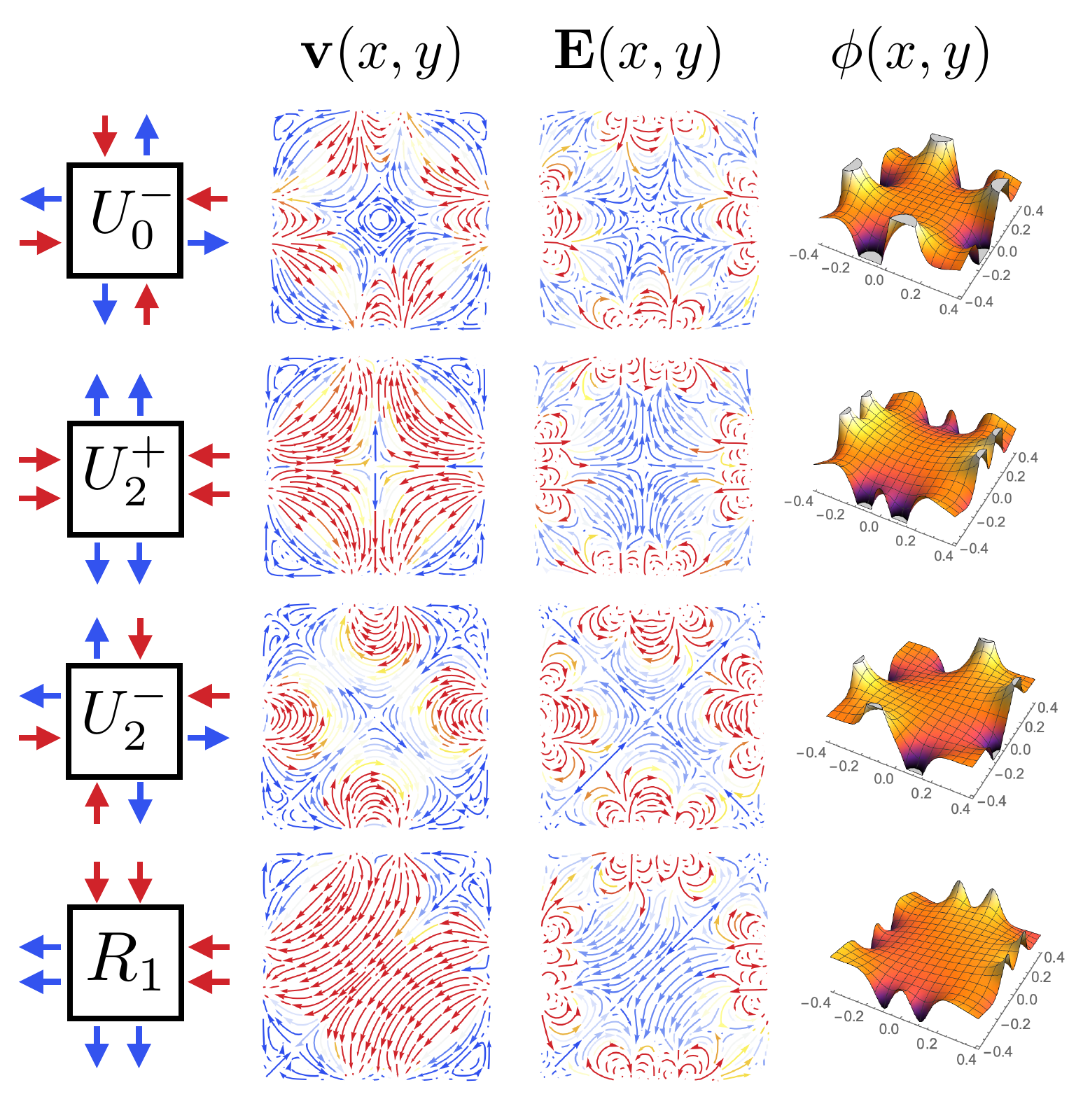}
    \caption{
    Viscous flows $\mathbf{v}$ and accompanying electric potentials $\phi$ and electric fields $\mathbf{E}=-\nabla{\phi}$ for various configurations of the dihedral viscometer. In the vector field plots for $\mathbf{v}$ and $\mathbf{E}$, color indicates vector magnitude, with red/blue indicating larger/smaller vectors. Parameter values $\delta=0$, $\lambda/w=\infty$, $d/w=0.20$, and $a/w=0.05$ taken in all plots. 
    }
    \label{fig:viscEfields}
\end{figure}

In this appendix, we present additional figures that summarize nice properties of our proposed viscometer for an experiment.   In \cref{fig:invertHeatsOHM} we demonstrate that the determination of $\mathsf{D}_8$ anisotropy $\delta$ is not substantially modified by momentum-relaxing scattering.  In \cref{fig:ohmSignal}, we further  demonstrate that the center heat signal is extremely robust to nonzero momentum relaxation, within a typical hydrodynamic regime $w\lesssim 5\lambda$. Even toward the ohmic limit at still stronger momentum-relaxation, only the rotational center heating is significantly affected.  \cref{fig:d2d} demonstrates a method to uniquely determine $\delta$ using only four total center heat measurements. \cref{fig:viscEfields} shows how the electric potentials and electric fields are expected to look for various configurations of the dihedral viscometer.

\label{sec:experiments}

\section{Flow plots}
\label{app:flowPlots}

\begin{figure}
\centering
\begin{subfigure}{0.4\textwidth}
    \centering
    \includegraphics[width=0.8\textwidth]{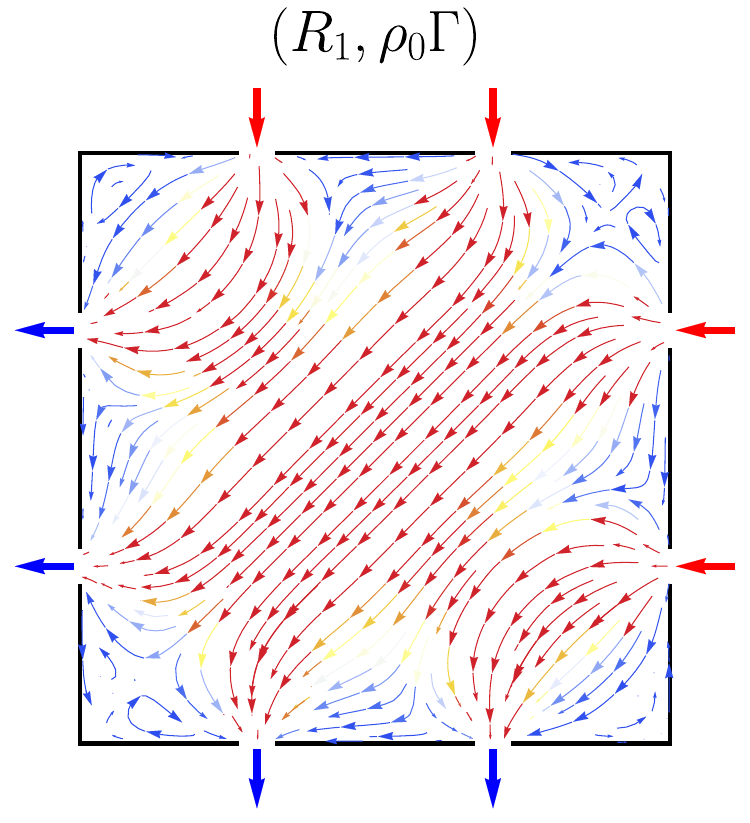}
    \caption{
    Configuration of our viscometer with boundary conditions of $R_1$ symmetry. The vector character of these boundary conditions preclude any viscous heating at the center of the viscometer, but do instead allow for a nonzero fluid flow $v_i\neq 0$ and hence nonzero ohmic heating $\rho_0\Gamma v_i^2\neq 0$ at the center.
    }
    \label{fig:sq8_ohmic}
\end{subfigure}
\;
\begin{subfigure}{0.4\textwidth}
    \centering
    \includegraphics[width=0.8\textwidth]{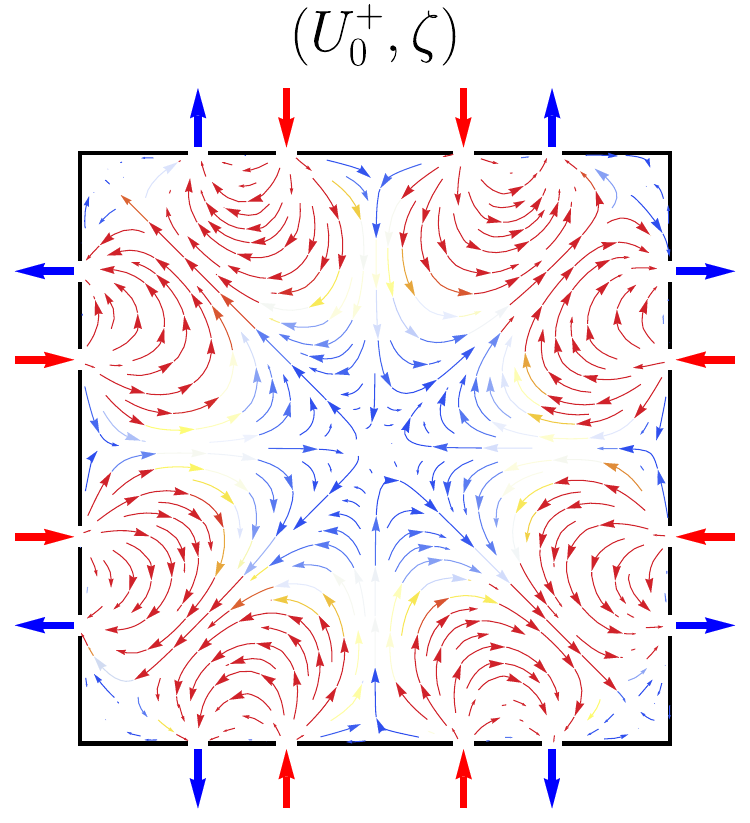}
    \caption{
    16-contact version of our dihedral viscometer with boundary conditions of $U_0^+$ symmetry. Although the total center heat for this case is mathematically zero in our incompressible approximation $v_i\approx\epsilon_{ij}\partial_j\psi$, boundary conditions of $U_0^+$ symmetry guarantee that the only possible hydrodynamic heat at the center can come from bulk viscous dissipation $\zeta$.
    }
    \label{fig:sq16_bulk}
\end{subfigure}
    \caption{
    Flows with boundary conditions transforming according to two remaining irreps of $\mathsf{D}_8$ not shown in \cref{fig:heatTable}. These irreps are labeled alongside the dissipative coefficient whose heat generation is isolated at the square center. Flow colors indicate the squared speed $\mathbf{v}^2$, with red representing higher speed and blue lower.
    }
    \label{fig:7}
\end{figure}

\begin{figure}[t]
\centering
\begin{subfigure}{\textwidth}
    \centering
    \includegraphics[width=0.9\textwidth]{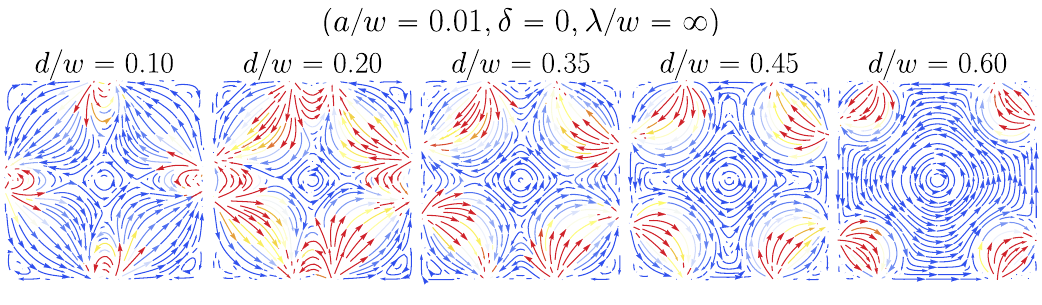}
    \caption{
    Varying contact spacing $d$.
    }
    \label{fig:rotD}
\end{subfigure}
\\
\begin{subfigure}{\textwidth}
    \centering
    \includegraphics[width=0.9\textwidth]{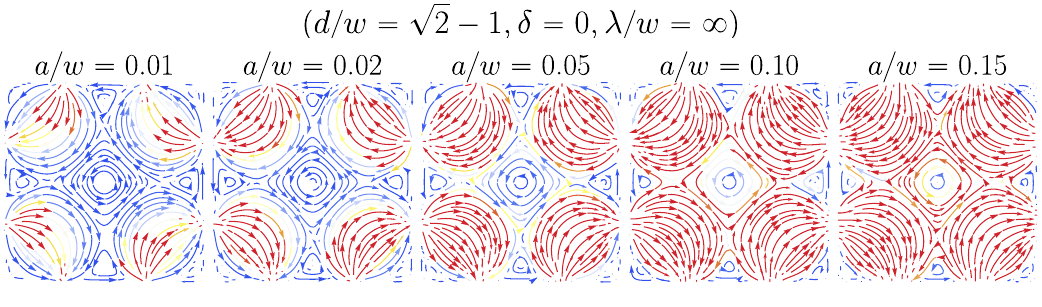}
    \caption{
    Varying contact width $a$.
    }
    \label{fig:rotA}
\end{subfigure}
\\
\begin{subfigure}{\textwidth}
    \centering
    \includegraphics[width=0.9\textwidth]{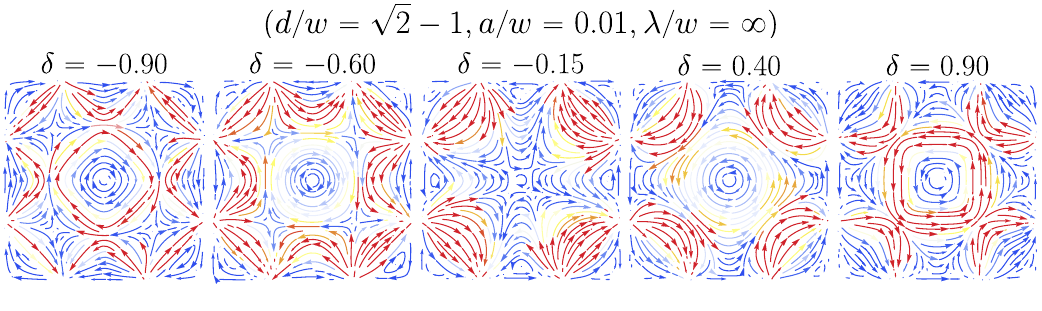}
    \caption{
    Varying $\mathsf{D}_8$ anisotropy $\delta$.
    }
    \label{fig:rotDelta}
\end{subfigure}
\\
\begin{subfigure}{\textwidth}
    \centering
    \includegraphics[width=0.9\textwidth]{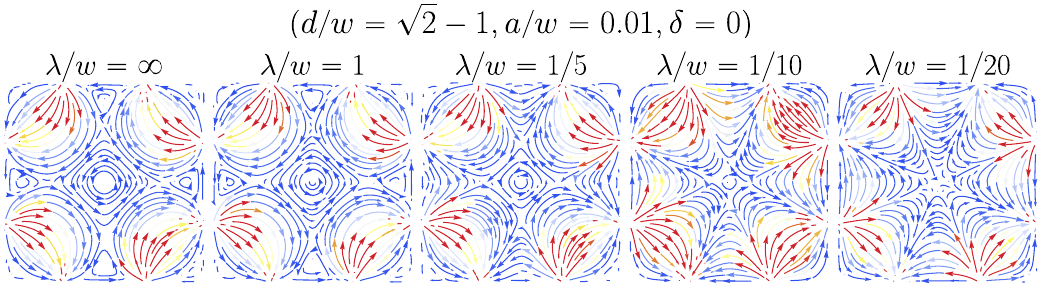}
    \caption{
    Varying Gurzhi length $\lambda$.
    }
    \label{fig:rotL}
\end{subfigure}
    \caption{
    Viscous flows in the dihedral viscometer in its $U_0^-$ configuration. 
    }
    \label{fig:8}
\end{figure}

\begin{figure}
\centering
\begin{subfigure}{\textwidth}
    \centering
    \includegraphics[width=0.9\textwidth]{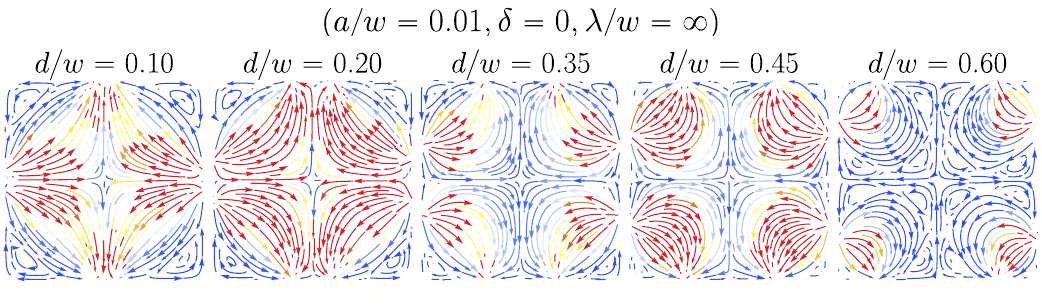}
    \caption{
    Varying contact spacing $d$.
    }
    \label{fig:plusD}
\end{subfigure}
\\
\begin{subfigure}{\textwidth}
    \centering
    \includegraphics[width=0.9\textwidth]{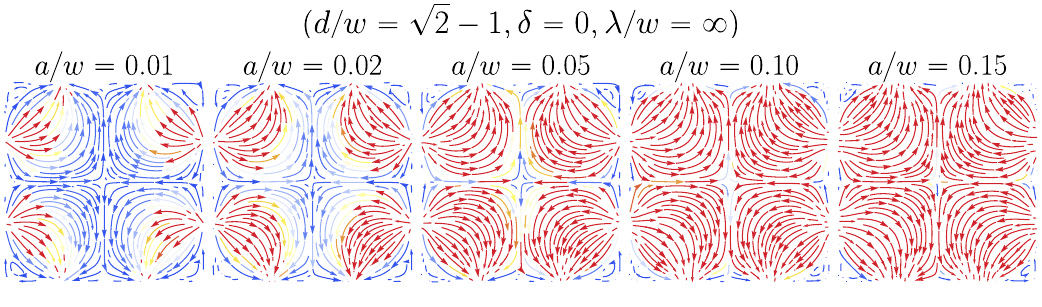}
    \caption{
    Varying contact width $a$.
    }
    \label{fig:plusA}
\end{subfigure}
\\
\begin{subfigure}{\textwidth}
    \centering
    \includegraphics[width=0.9\textwidth]{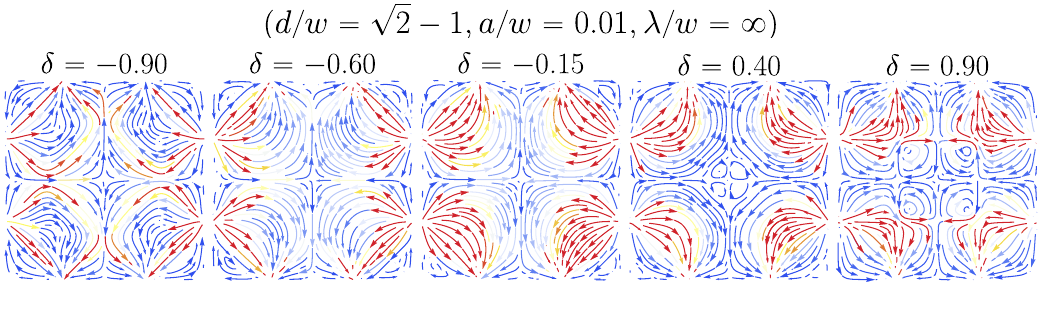}
    \caption{
    Varying $\mathsf{D}_8$ anisotropy $\delta$.
    }
    \label{fig:plusDelta}
\end{subfigure}
\\
\begin{subfigure}{\textwidth}
    \centering
    \includegraphics[width=0.9\textwidth]{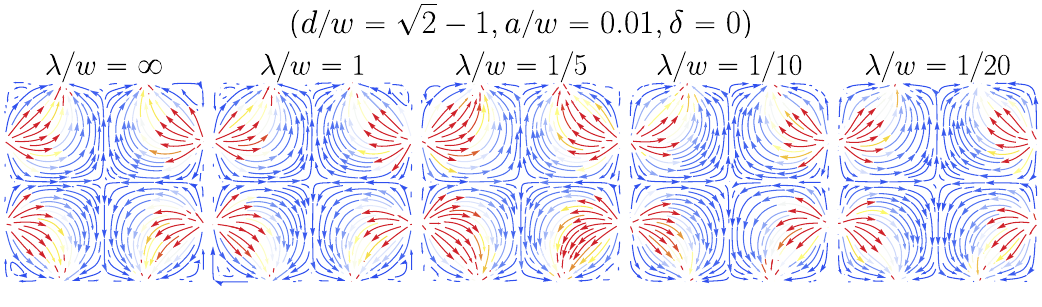}
    \caption{
    Varying Gurzhi length $\lambda$.
    }
    \label{fig:plusL}
\end{subfigure}
    \caption{
    Viscous flows in the dihedral viscometer in its $U_2^+$ configuration. 
    }
    \label{fig:9}
\end{figure}

\begin{figure}
\centering
\begin{subfigure}{\textwidth}
    \centering
    \includegraphics[width=0.9\textwidth]{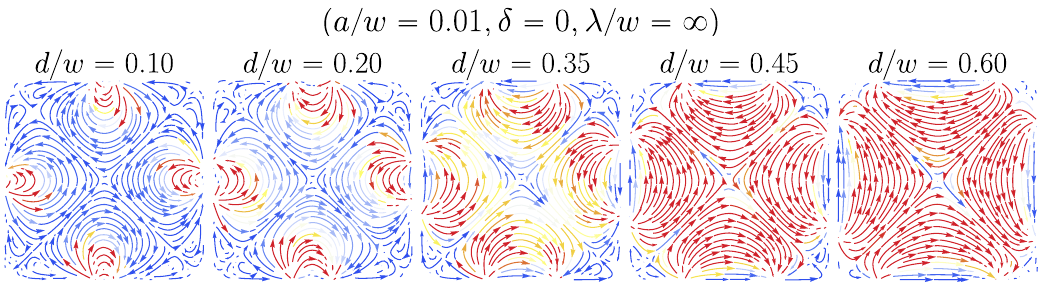}
    \caption{
    Varying contact spacing $d$.
    }
    \label{fig:timesD}
\end{subfigure}
\\
\begin{subfigure}{\textwidth}
    \centering
    \includegraphics[width=0.9\textwidth]{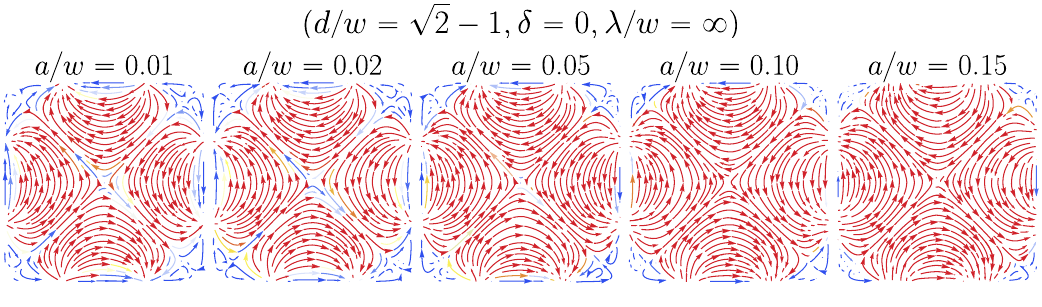}
    \caption{
    Varying contact width $a$.
    }
    \label{fig:timesA}
\end{subfigure}
\\
\begin{subfigure}{\textwidth}
    \centering
    \includegraphics[width=0.9\textwidth]{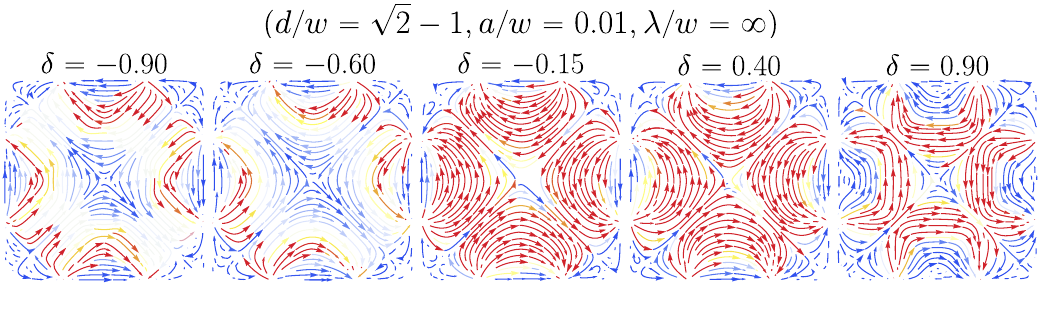}
    \caption{
    Varying $\mathsf{D}_8$ anisotropy $\delta$.
    }
    \label{fig:timesDelta}
\end{subfigure}
\\
\begin{subfigure}{\textwidth}
    \centering
    \includegraphics[width=0.9\textwidth]{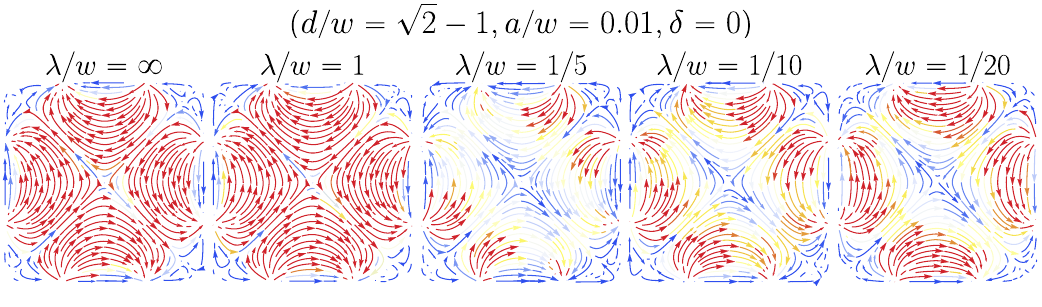}
    \caption{
    Varying Gurzhi length $\lambda$.
    }
    \label{fig:timesL}
\end{subfigure}
    \caption{
    Viscous flows in the dihedral viscometer in its $U_2^-$ configuration. 
    }
    \label{fig:10}
\end{figure}

\begin{figure}
    \centering
    \includegraphics[width=.85\textwidth]{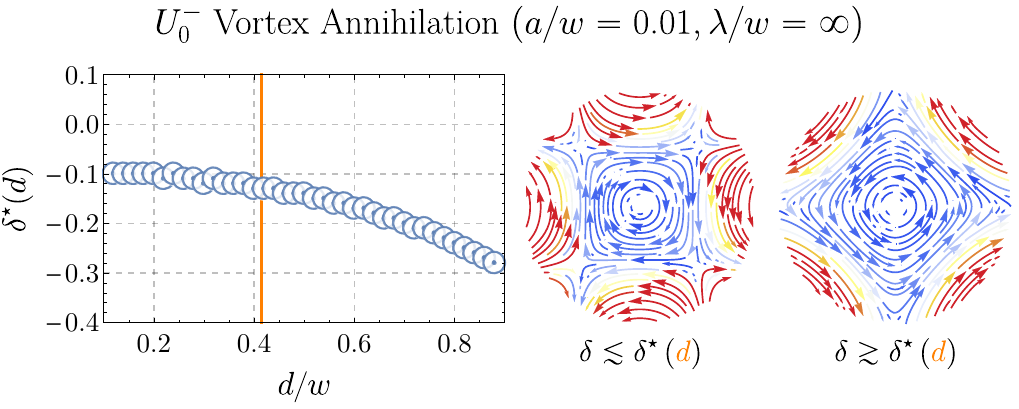}
    \caption{\textit{Left}: Numerical estimates of the critical $\mathsf{D}_8$ anisotropy $\delta^\star$, across which the the central $U_0^-$ vortex closes and re-opens (rotated $45^\circ$ and with opposite vorticity), as a function of the contact spacing $d$. \textit{Right}: A zoomed-in view of the central $U_0^-$ vortex for $d/w=(\sqrt{2}-1)\approx0.41$, just below and above the transition.
    }
    \label{fig:vortexDestruct}
\end{figure}

\begin{figure}
    \centering
    \includegraphics[width=.85\textwidth]{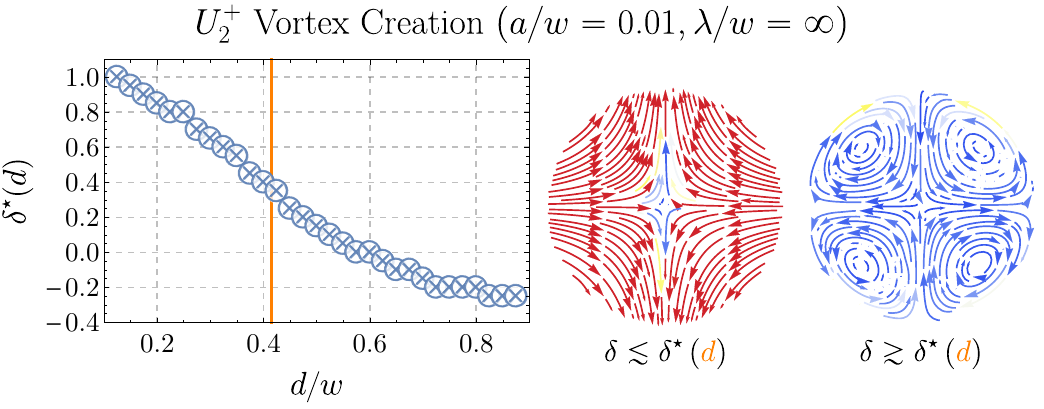}
    \caption{\textit{Left}: Numerical estimates of the critical $\mathsf{D}_8$ anisotropy $\delta^\star$, across which the the $U_2^+$ center becomes unstable to fourfold vortex production, as a function of the contact spacing $d$. \textit{Right}: A zoomed-in view of the the $U_2^+$ center for $d/w=(\sqrt{2}-1)\approx0.41$, just below and above the transition.
    }
    \label{fig:vortexDestruct2}
\end{figure}

In this appendix, we collect some useful plots that demonstrate flow patterns in our proposed viscometer, including how they change as a function of parameters.  \cref{fig:7} shows the $R_1$ and $U_0^+$ flow patterns that we did not show in the main text. \cref{fig:8} shows a diversity of flow patterns in the $U_0^-$ configuration; \cref{fig:9} in the $U_2^+$ configuration; and \cref{fig:10} in the $U_2^-$ configuration.  \cref{fig:vortexDestruct} shows how the rotational viscosity signal disappears as a function of $\delta$ as the center vortex switches orientation; \cref{fig:vortexDestruct2} shows the formation of 4 vortices at the center of the viscometer in the $U_2^+$ configuration. 

\end{appendix}
\twocolumngrid

\end{document}